\documentstyle[11pt,aaspp4]{article}

\lefthead{Trotter, Winn \& Hewitt}
\righthead{
Models of the gravitational lens MG~J0414+0534
}

\def\Del{{\mbox{\boldmath{$\nabla$}}}}
\def\lDel{{\mbox{\boldmath{$\nabla$}}}}
\def\svec{{\bf s}}
\def\rvec{{\bf r}}
\def\xvec{{\bf x}}
\def\cvec{{\bf c}}
\def\sigvec{{\mbox{\boldmath $\sigma$}}}
\def\Phivec{{{\bf \Phi}}}
\def\Bvec{{{\bf B}}}
\def\Avec{{{\bf A}}}
\def\Fvec{{{\bf F}}}
\def\Gvec{{{\bf G}}}
\def\Mvec{{{\bf M}}}
\def\xhat{{\hat{\bf x}}}
\def\yhat{{\hat{\bf y}}}
\def\rhat{{\hat{\bf r}}}
\def\thetahat{{\hat{\mbox{\boldmath $\theta$}}}}

\def\sigcrit{{\sigma_{\rm c}}}
\def\mth{{$m^{\rm th}$}}

\def\mc{{m{\rm c}}}
\def\ms{{m{\rm s}}}
\def\mt{{mt}}
\def\mtc{{mt{\rm c}}}
\def\mts{{mt{\rm s}}}

\def\orderof{{\cal O}}

\newcommand{\Sumfrom}[1]{\sum_{#1}^{\infty}}

\def\intztp{{\int_0^{2\pi}}}
\def\massDensAngM{{\psi_{\sigma_m}}}
\def\ehatmlongparen{{({{\xhat \cos m\theta + \yhat \sin m\theta}})}}
\def\Phimnearring{\Phivec_{m}^{\rm near \,  ring}}

\def\Bmvec{\Bvec_m}
\def\Bmc{B_\mc}
\def\Bms{B_\ms}

\newcommand{\BmvecWith}[1]{\Bvec_{#1}}
\newcommand{\BmAmpWith}[1]{B_{#1}}
\newcommand{\BmAngWith}[1]{\psi_{B_{#1}}}
\def\Amvec{\Avec_m}
\def\Amc{A_\mc}
\def\Ams{A_\ms}
\def\AmAmp{A_m}

\newcommand{\AmvecWith}[1]{\Avec_{#1}}
\newcommand{\AmAmpWith}[1]{A_{#1}}
\newcommand{\AmAngWith}[1]{\psi_{A_{#1}}}
\def\bE{b}		
\def\ft{f_t}  	
\newcommand{\ftWith}[1]{f_{#1}}	
\def\Fmtvec{\Fvec_\mt}
\def\Mdiffmvec{{\Mvec^{\rm diff}_{m}}}	
\newcommand{\MdiffmvecWith}[1]{\Mvec^{\rm diff}_{#1}}
\def\Mdiffmc{{M^{\rm diff}_{\mc}}}		
\def\Mdiffms{{M^{\rm diff}_{\ms}}}		
\def\Msummvec{{\Mvec^{\rm sum}_{m}}}
\newcommand{\MsummvecWith}[1]{{\Mvec^{\rm sum}_{#1}}}
\newcommand{\MsummvecTab}[1]{{\Mvec^{\rm sum}_{#1}}}
\def\Msummc{{M^{\rm sum}_{\mc}}}		
\def\Msumms{{M^{\rm sum}_{\ms}}}		
\def\Gmtvec{\Gvec_\mt}
\def\Gmtc{G_\mtc}	
\def\Gmts{G_\mts}	

\def\zerothderivOneD{}
\def\firstderivOneD{{\bE \frac{d}{dr}}} 
\def\secondderivOneD{{\bE^2 \frac{d^2}{dr^2}}}
\def\thirdderivOneD{{\bE^3 \frac{d^3}{dr^3}}}
\def\tthderivOneD{{\bE^t \frac{d^t}{dr^t}}}
\def\zerothderivOneDOverBSqr{{\frac{1}{\bE^2} \zerothderivOneD}}
\def\firstderivOneDOverBSqr{{\frac{1}{\bE^2}  \firstderivOneD}}
\def\secondderivOneDOverBSqr{{\frac{1}{\bE^2} \secondderivOneD}}
\def\thirdderivOneDOverBSqr{{\frac{1}{\bE^2}  \thirdderivOneD}}
\def\tthderivOneDOverBSqr{{\frac{1}{\bE^2}    \tthderivOneD}}
\def\rescale{{(1-\kappa)}}
\def\vstrut{\rule{0pt}{4.5ex}}
\def\leftParen{\left[}
\def\rightParen{\right]}
\def\Msumone{{ \Msummc' \cos m \theta + \Msumms' \sin m \theta }}
\def\Msumtwo{{ \Msummc' \sin m \theta - \Msumms' \cos m \theta }}
\def\Mdiffone{{ \Mdiffmc' \cos m \theta + \Mdiffms' \sin m \theta }}
\def\Mdifftwo{{ \Mdiffmc' \sin m \theta - \Mdiffms' \cos m \theta }}
\def\Aone{{ \Amc' \cos m \theta + \Ams' \sin m \theta }}
\def\Atwo{{ \Amc' \sin m \theta - \Ams' \cos m \theta }}
\def\Bone{{ \Bmc' \cos m \theta + \Bms' \sin m \theta }}
\def\Btwo{{ \Bmc' \sin m \theta - \Bms' \cos m \theta }}
\def\Gone{{ \Gmtc' \cos m \theta + \Gmts' \sin m \theta }}
\def\Gtwo{{ \Gmtc' \sin m \theta - \Gmts' \cos m \theta }}

\def\tstrut{\rule{0pt}{2.2ex}}

\def\formalerrors{{\renewcommand{\arraystretch}{0}\begin{array}[t]{@{}l@{}} {\scriptscriptstyle\rm formal} \\ {\scriptscriptstyle\rm errors \rule{0cm}{1ex}} \end{array}}}   
\def\whichmodel{{\renewcommand{\arraystretch}{0}\begin{array}[t]{@{}l@{}} {\scriptscriptstyle\rm which} \\ {\scriptscriptstyle\rm model \rule{0cm}{1ex}} \end{array}}}     
\def\AoneAtwodifference{{\renewcommand{\arraystretch}{0}\begin{array}[t]{@{}l@{}} {\scriptscriptstyle\rm A1-A2} \\ {\scriptscriptstyle\rm difference \rule{0cm}{1ex}} \end{array}}}

\begin{document}

\title{
A multipole-Taylor expansion for the potential \\
of gravitational lens MG~J0414+0534
}

\author{
Catherine S. Trotter\footnote[1]{cst@alum.mit.edu},
Joshua N. Winn\footnote[2]{jnwinn@mit.edu}
and Jacqueline N. Hewitt\footnote[3]{jhewitt@maggie.mit.edu}
}

\affil{Department of Physics, Massachusetts Institute of Technology,
Cambridge, MA 02139}

\begin{abstract}

We employ a multipole-Taylor expansion to
investigate how tightly the gravitational potential
of the quadruple-image lens MG~J0414+0534 is
constrained by recent VLBI observations.
These observations revealed that each of
the four images of the background radio source contains four distinct
components, thereby providing more numerous and more precise constraints
on the lens potential than were previously available.
We expand the
two-dimensional lens potential
using multipoles for the angular coordinate and a modified
Taylor series for the radial coordinate.
After discussing the physical significance of each term,
we compute models of MG~J0414+0534 using only
VLBI positions as constraints.
The best-fit model has both interior and exterior quadrupole moments
as well as exterior $m=3$ and $m=4$ multipole moments. The deflector centroid
in the models matches the optical galaxy position,
and the quadrupoles are aligned
with the optical isophotes. The radial distribution of mass could not
be well constrained. We discuss the implications of these models for the
deflector mass distribution and for
the predicted time delays between lensed components.

\end{abstract}

\keywords{gravitational lensing --- methods: data analysis --- galaxies: structure}

\section{Introduction}

There are now about 40 observed examples
of ``strong'' gravitational lensing,
in which multiple images of a background object are produced by the
gravitational potential of an intervening object (\cite{castles}).
Constructing models for the gravitational potential of the deflecting
mass in these systems is of fundamental importance, in the first place
to verify that the observed system is truly a gravitational
lens. A crucial hurdle for every promising gravitational lens candidate
is to have its image configuration reproduced, at least approximately, by
a plausible model for the deflecting mass
distribution. If this cannot be achieved, the lensing interpretation must
be seriously doubted.

Once this and other tests for
lensing are passed successfully, however, model construction moves
beyond a plausibility check into a direct measurement of the
mass distribution of the lens.
Lensing makes a unique contribution because it 
is sensitive to all forms of mass (including dark matter), yet 
does not depend on any luminous tracers in the lens.
Lens modeling is also a crucial step in the enterprise of determining
cosmological parameters by measuring the time delays between
the light curves of multiple images.
A successful model will predict the values of these time delays in terms
of parameters such as $H_{0}$, $\Omega_{m}$, and $\Omega_{\Lambda}$.
These parameters can then be constrained by the
time-delay measurements (\cite{refsdal};
for recent examples see \cite{delay1}, \cite{delay2}). The appeal of
this technique is that it does not make use of the usual chain
of intermediate distance indicators and their associated uncertainties.

The determinations of mass distributions and cosmological parameters
are both frustrated by the main challenge of lens modeling: it
is a poorly-constrained inverse problem with unknown systematic errors.
It is neither obvious which parameterized model for the
gravitational potential should be chosen, nor how precisely the parameters
are constrained by the data.
The choice of parameterization for the gravitational potential is usually
dictated by two factors: preconceptions about the mass distribution
of galaxies (based on the distribution of luminous matter and/or velocity
dispersions), and ease of computation. Some examples are singular
isothermal spheres, elliptical density profiles, elliptical potential
contours, and triaxial density profiles (see e.g. \cite{model1};
\cite{model2}; \cite{model3}). Added to these are
terms representing external ``shear'' from the tidal forces of
neighboring mass concentrations. Such models have all been
used successfully to reproduce
the image configurations of the known sample of lenses,
at least qualitatively,
which leads one to wonder how well the observations of strong
lensing actually constrain the lens potential.

Kochanek (1991) was the first to investigate
this question systematically.
He attempted to model each of 12 different lenses using a set of 6 different
parameterized potentials. The potentials he considered were point masses
and singular isothermal spheres, added to shear terms with one of three
radial dependencies. He found that the total mass in the region between the
multiple images is well constrained, but the radial dependence of the
potential is not. This was because the multiple images are
typically located close to the ``Einstein ring'' of the lens, so the
constraints on the potential are correspondingly limited to a small range
in radius.
Kochanek suggested (but did not carry out) a Taylor expansion, parameterized
by the deviation from the ring
radius, as a way to turn this fact to advantage.

The purpose of this paper is to elaborate upon this suggestion and apply it
to a particular quadruple-image gravitational lens, MG~J0414+0534. We expand the 
potential as a series of multipoles in angle, and 
as a modified Taylor series in radius.
The resulting series has three
advantages over traditional lens models.  One, it is mathematically
general, and therefore less subject to
preconceptions about the galactic mass distribution
(which, after all, is what is trying to be determined).
Two, the degeneracies in the lens equation and the moments of
the mass distribution have simple correspondences to
individual terms of the series.
Three, each successive term in the series is expected to contribute a
smaller amount to the light deflection, so long as the radial parameter is
small and the angular dependence of the potential is
smooth. This allows the complexity of the model to be easily prescribed
by the truncation of the series. However, although the first assumption
(the smallness of the radial parameter) is empirically true for
quadruple-image lenses, the second assumption (the angular smoothness of the
potential) is open to question. 
Nevertheless, in the
face of our ignorance of true galaxy structure, we employ a multipole expansion
because of its simplicity and generality.

The disadvantage of a mathematically general expansion (rather than one that
is motivated by astrophysical preconceptions) is that the
number of parameters is large. In many cases, especially for the double-image
lenses, the number of parameters would
be comparable to
or more than the number of observational constraints. This is probably
why such an expansion has not been used previously.
However, recent VLBI images
of MG~J0414+0534 provide a much larger
body of constraints than were previously available (\cite{cathythesis}). Each of the
four lensed images was found to contain four components, with clear
correspondences between the components of different images.  This provides 16 
components whose positions are known with milliarcsecond precision,
thereby creating a testbed for the multipole-Taylor
expansion.


This paper will be organized as follows. The next section describes
prior optical and radio observations of MG~J0414+0534, in addition to
the latest VLBI map.  Section~\ref{sec:expansion} 
presents the formalism for our series
expansion, discusses the physical significance of each term, and identifies
the terms that cannot be constrained due to degeneracies in the lens equation.
Some previous models for MG~J0414+0534 are discussed and compared to our
technique. Section~\ref{sec:application}
explains the numerical methods we employed and 
presents the best-fit results.  Finally,
section~\ref{sec:discussion} discusses the implications of these
results for the mass distribution of the lens and the time delays
between images. 

\section{Observations of MG~J0414+0534}
\label{sec:observations}

The radio source MG~J0414+0534 was first identified by
Hewitt et al. (1992) as a gravitational
lens during a systematic search for lenses in the MIT-Green
Bank 5 GHz radio catalog. In both optical and radio
images it has four components, which have been called A1, A2, B, and C,
in order of decreasing brightness.
The radio components all have spectral index\footnote[1]{The spectral index $\alpha$
is defined such that
$S_{\nu} \propto \nu^{\alpha}$, where $S_{\nu}$ is the spectral flux density.}
$\alpha = -0.80 \pm 0.02$
(\cite{katz}),
and the optical components are all exceedingly red. However, the A1/A2
radio flux ratio ($\sim 1.1$) and optical flux ratio ($\sim 2.5$)
do not agree, even though gravitational light deflection is
independent of wavelength. The discrepancy could be caused by a
number of factors (\cite{angonin}), including dust (\cite{dust})
and microlensing (\cite{wittmao}).

Schechter \& Moore (1993) discovered the lensing galaxy in the $I$ band, along
with a faint object $1\arcsec$
west of component B which they named object X. The optical/infrared
spectrum of the lensed source resembles a very reddened quasar at
$z_s = 2.639 \pm 0.002$ (\cite{sourcez}). The redshift of the lensing
galaxy is $z_l = 0.9584 \pm 0.0002$ (\cite{lensz}).
The best presently published optical photometry (uncertainty $\approx 0.004$ mag)
and astrometry (uncertainty $\approx 0.02 \arcsec$)
for this system were obtained with Hubble Space
Telescope (HST) WFPC2/PC1 observations by \cite{hst},
who also detected a blue arc extending from A1 to A2 to B.

The deepest radio observations to date did not detect a fifth
radio component in the system (\cite{katz}).  Higher-resolution radio
maps, using MERLIN (\cite{merlin}) and VLBI (\cite{patnaik}),
showed substructure
within the four images of MG~J0414+0534.
In particular, the VLBI map
resolved images A1, A2, and B into two components each, and
revealed image C to be extended.
Previous attempts to use VLBI
constraints to perform lens modeling, however, were hampered
by the low signal-to-noise ratio of the available data (\cite{ellithorpe}).

\begin{figure}
\plotone{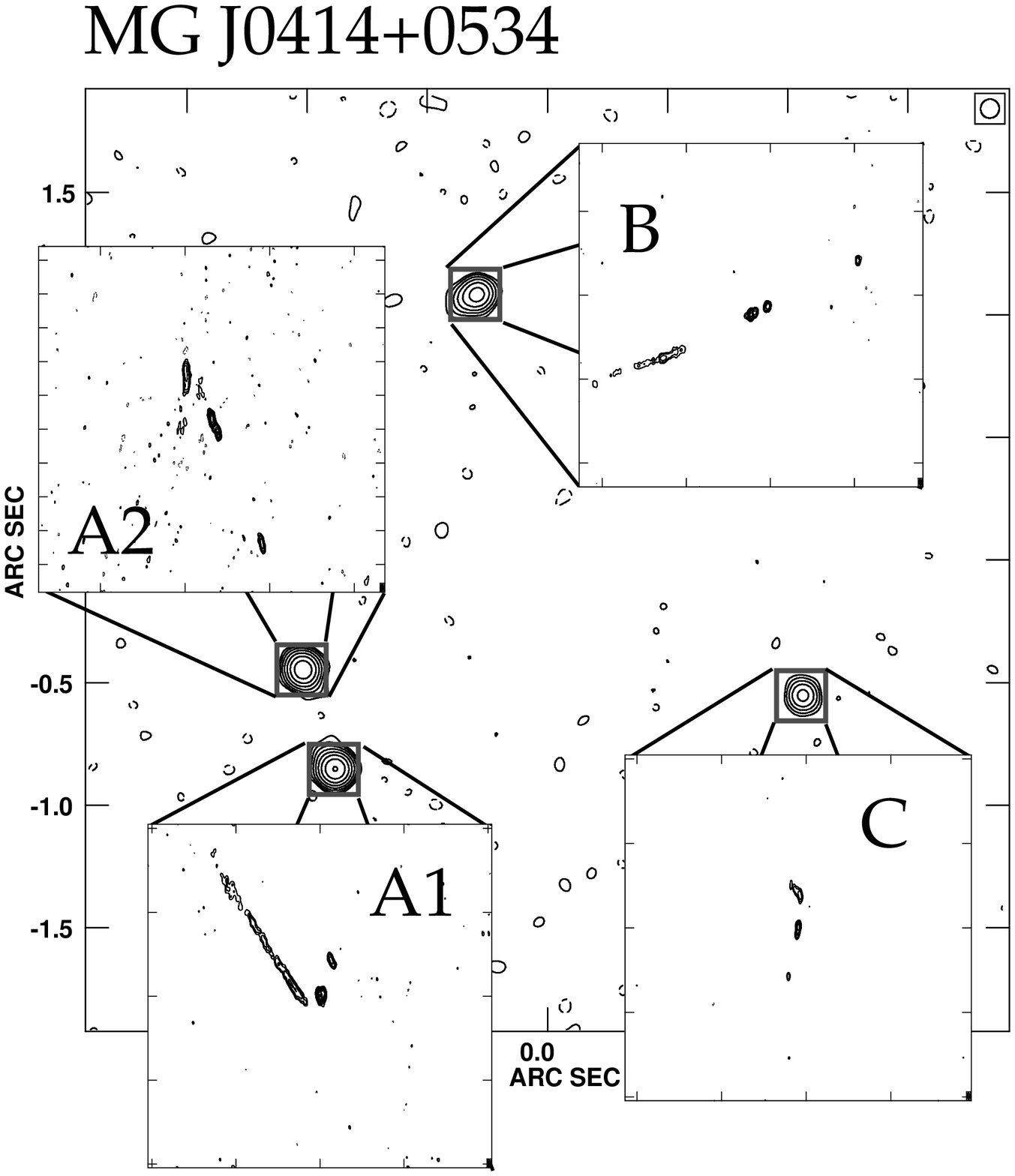}
\caption{High-resolution maps of the four images of MG~J0414+0534,
observed at 5 GHz with the VLBA, are shown superimposed on the
22 GHz VLA map of \protect\cite{katz} (kindly supplied by C. Katz).} 
\label{fig:observation}
\end{figure}

\begin{figure}
\plotone{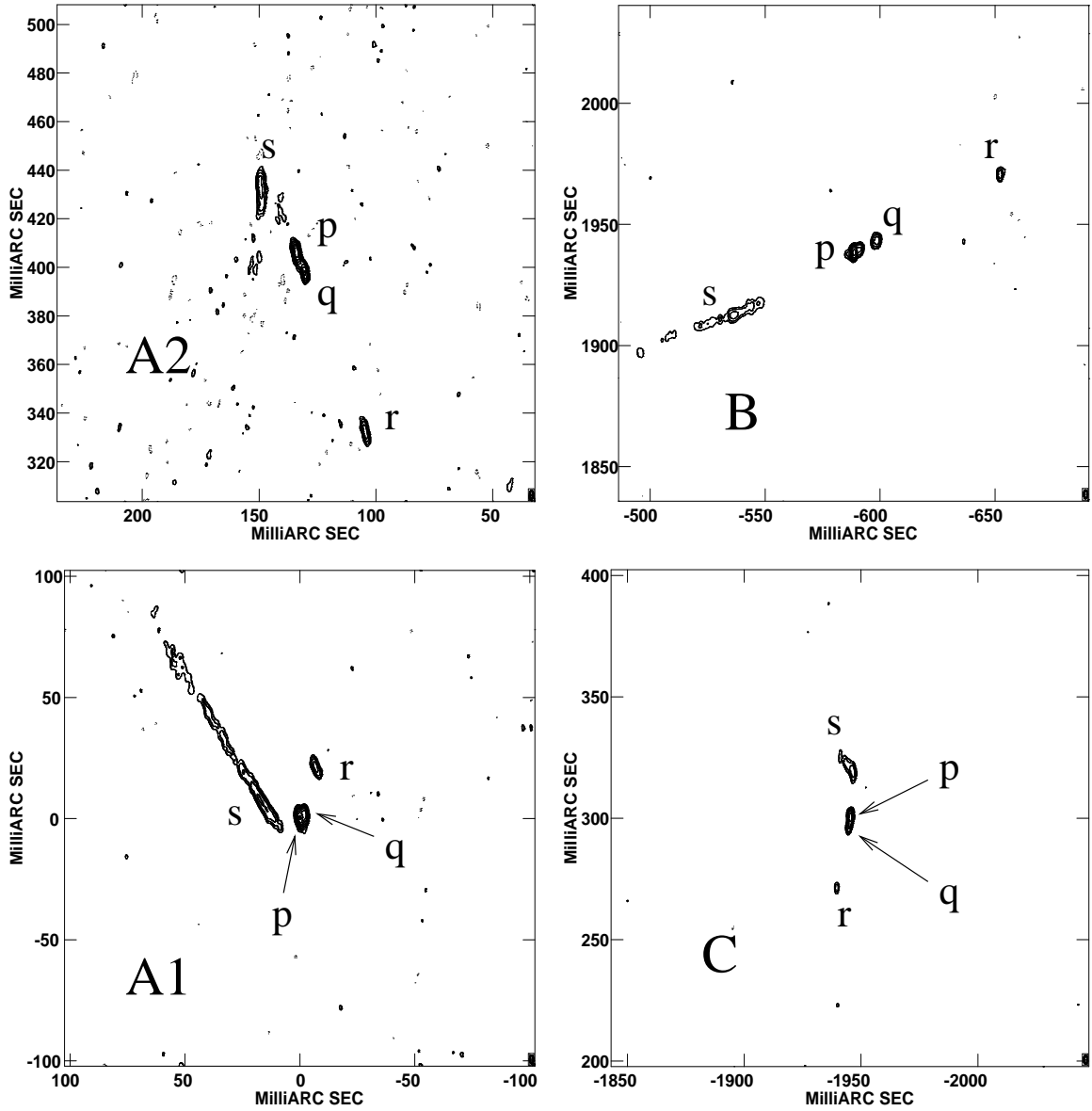}
\caption{
High-resolution maps of the four images of MG~J0414+0534,
observed at 5 GHz with the VLBA, are shown with the four
components p, q, r and s identified. The positions and fluxes
for these components are reported in Table~\ref{tbl:positions}.
}
\label{fig:bigmaps}
\end{figure}

As part of a 1995 study to test the feasibility of using the
NRAO\footnote[2]{The National
Radio Astronomy Observatory (NRAO)
is operated by Associated Universities, Inc.,
under cooperative agreement with the National Science Foundation.}
Very Long Baseline Array (VLBA) 
to monitor the time-variability of various radio-loud lenses, we obtained
5~GHz images of MG~J0414+0534. These observations and the data reduction
procedure are described in detail elsewhere (\cite{cathythesis}) and will
be presented in a future paper; only the results are summarized here.
The synthesized beam was $1.5 \times 3.5$ milliarcseconds,
and the RMS thermal noise was about 0.15 mJy/beam, close to the theoretical
limit. The peak fluxes in the maps of each of the four images ranged
from 21 to 110 mJy/beam, allowing a much higher signal-to-noise ratio
than was previously available. Within each image were detected
four components. The compact components
p, q, and r were labeled in order
of decreasing brightness; there is also one extended component,
labeled s. The four maps are shown in Figure~\ref{fig:observation}, 
superimposed on
a lower-resolution VLA map.  The sub-components are labeled in the
larger maps shown in Figure~\ref{fig:bigmaps}.
The locations, fluxes, and extents of
the components were determined by least-squares fitting to a set of
elliptical Gaussian parameters. Table~\ref{tbl:positions}
lists the locations and fluxes of the 16 VLBI components.

\section{The modified multipole-Taylor expansion}
\label{sec:expansion}

The motivation behind our method for lens modeling
is to expand the lens potential
in a manner which preserves a measure of mathematical generality
and in which the terms that can and cannot be constrained by lensing
data are clearly separated. We have tried to arrange for the
parameters in our expansion to be sensitive to what can be learned from
lensing data, rather than what one would wish {\em a priori} to learn.
In this section we will develop this expansion in detail,
describe the physical significance and constrainability of each term,
and compare our expansion to other commonly-used parameterizations.

\subsection{The multipole-Taylor expansion}

The goal of lens modeling is to deduce the
two-dimensional gravitational lens potential, which is defined in a way
similar to the definitions of
Schneider, Ehlers \& Falco (1992) and Narayan \& Bartelmann (1996):
\begin{equation}
\Phi(\rvec) = \frac{2D_{LS}}{D_{L}D_{S}} \int{dz\ \Phi_{N}(D_{L}\rvec,z)}.
\end{equation}
Here $\rvec$ is the angular coordinate measured from an arbitrary
optic axis, $z$ is the line-of-sight coordinate, and $\Phi_{N}$ is the Newtonian
gravitational potential of the lens. 
The angular diameter distances $D_{L}$, $D_{S}$, and $D_{LS}$ are from
observer to lens, observer to source, and lens to source, respectively.
As elsewhere in this paper, the speed of light has been set to 1 by a
choice of units.

The lens potential is related to the surface mass density of the lensing
object by the two-dimensional Poisson equation,

\begin{equation}
\label{eq:poisson-equation}
\nabla_{\rvec}^2 \Phi(\rvec) = 
8\pi G \frac{D_{L}D_{LS}}{D_{S}} \sigma(\rvec) = 
\frac{2\sigma(\rvec)}{\sigcrit}.
\end{equation}
Implicit in the above equation is the definition of the critical surface mass density,
$\sigcrit = (1/4\pi G)(D_{S}/D_{L}D_{LS})$. The lens potential
determines the image configuration via the lens equation,
\begin{equation}
\label{eq:lens-equation}
\svec = \rvec - \Del_{\rvec}\Phi(\rvec),
\end{equation}
where $\rvec$ is the image position and $\svec$ is the source position.
The typical procedure in lens modeling is to adopt a parameterized form
for $\Phi$, and then fix the parameters by minimizing an error function
$\chi^{2}$. The error function may represent the deviations between the observed
images and the images that are projected through the model potential from
a model source distribution. Alternatively, the error function may
represent the deviations between source positions that are back-projected
from the various images (which is computationally simpler), as explained further
in section~\ref{sec:source-plane-chi}.

The parameterization we adopt is a multipole-Taylor expansion.
The first step is to expand the potential
$\Phi(\rvec)$ 
in terms of a complete set of orthogonal basis
functions which make the Poisson equation separable:
\begin{equation}
\label{eq:multipole-series}
\Phi(r,\theta) = \Phi_0(r) + \sum_{m=1}^{\infty} \Phivec_m(r) \cdot 
( \xhat \cos m\theta + \yhat \sin m\theta )
\end{equation}
The function $\Phi_{0}(r)$ is the monopole, and the vector-like
functions $\Phivec_m(r)$ are the higher multipoles.
Schneider \& Weiss (1991) also carried out a multipole expansion of the lens
potential in this manner.
The motivation for this expansion is that each multipole moment
of the potential depends only on the corresponding multipole moment
of the surface mass density. Specifically, the relations are
(\cite{koch91}):
\begin{eqnarray}
\label{eq:poisson-monopole}
\Phi_{0}(r) &=& {\rm constant}
        + 2 \ln r \int_0^r dr' \, r' \frac{\sigma_0(r')}{\sigcrit} 
        + 2 \int_r^\infty dr' \, r' \ln r' \frac{\sigma_0(r')}{\sigcrit} 
                \\
\label{eq:poisson-multipole}
\Phivec_{m}(r) &=&
   -\frac{r^{-m}}{m} \int_0^r dr' \, r'^{(1+m)}
    \frac{\sigvec_{m}(r')}{\sigcrit} 
   -\frac{r^{ m}}{m} \int_r^\infty dr' \, r'^{(1-m)}
    \frac{\sigvec_{m}(r')}{\sigcrit} 
.
\end{eqnarray}

For smooth mass distributions, only the few lowest-order terms
in the multipole expansion of the surface mass density
are expected to be significant, and the rest may be neglected. 
By reference to equations~\ref{eq:poisson-monopole}
and~\ref{eq:poisson-multipole} it is clear that 
this is equivalent to neglecting the higher
multipole components of the potential.
For modeling purposes,
one may truncate the expansion of
the potential at the desired level.

The radial dependence of $\Phi_0(r)$ and $\Phivec_m(r)$
must also be parameterized to be suitable for use in lens modeling.
Lensed images constrain $\Del\Phi$ and
$\partial_i \partial_j \Phi$
at the locations of the images. For Einstein rings and quadruple-image lenses
(``quads''), the images are
typically near the lens's characteristic ring radius.
Therefore, following the suggestion of Kochanek (1991), we
expand the radial dependence of each multipole moment of the potential
as a Taylor series in the parameter $\rho = (r-b)/b$, where $b$ is the
Einstein ring radius. 
(The meaning of the Einstein ring radius for a non-circular lens 
is discussed below, in section~\ref{sec:significance}.)
This series should converge quickly for 
image positions located at $r \approx b$.
For example, $|\rho| < 0.2$ for all of the components in MG~J0414+0534.

The resulting expansion of the lens potential is:
\begin{eqnarray}
\label{eq:multipole-taylor-series}
\frac{1}{b^2}\Phi(r, \theta)
        &=& {\rm const} 
                + \rho 
                + \frac{1}{2} \rho^2 \ftWith{2}
                + \Sumfrom{t=3} \frac{1}{t!} \rho^t \ft
\\ & &
                - \Sumfrom{m=1} \left(
                \left\{ 
                        \Msummvec + \rho m \Mdiffmvec
                        + \Sumfrom{t=2} \frac{1}{t!} \rho^t \Fmtvec
                \right\} \cdot \ehatmlongparen
                \right)
, \nonumber 
\end{eqnarray}
parametrized by the origin of the expansion $(g_x, g_y)$, the ring
radius $b$, the monopole parameters $f_t$, and the higher
multipole parameters $\Msummvec$, $\Mdiffmvec$, and $\Fmtvec$.  
We use $t$ as an integral index,
contrary to convention, because of the mnemonic value of associating
$t$ with Taylor and $m$ with multipole.  
The additive constant $f_0$ does not affect light deflection.
Because we factor out the ring radius $b$,
and then use it as a parameter, we have $f_1 = 1$.
The $m\geq1$ multipoles with $t=0$ and $t=1$ received the
special names $\Msummvec$ and $\Mdiffmvec$ for reasons that will
become apparent in the next section.  
The model parameters $g_x$
and $g_y$, which specify the location of the origin of coordinates,
are implicit in equation~\ref{eq:multipole-taylor-series}.  To
simplify the interpretation of the multipole moments, this origin
should be centered on the deflector.  In
section~\ref{sec:degeneracies} it will be shown how this condition may
be enforced during the model-fitting procedure.

\subsection{Physical significance of the expansion parameters}
\label{sec:significance}

Although the potential $\Phi(\rvec)$ is the quantity most directly constrained by
observations of lensing, it is the surface mass density $\sigma(\rvec)$ that
is usually of direct astrophysical interest. 
In this section the correspondence between the 
the parameters in the multipole-Taylor expansion of $\Phi(\rvec)$ 
and the multipoles of $\sigma(\rvec)$ will be made explicit.
This is useful because the multipole moments of the surface mass
density have a simple physical meaning.  This meaning will be reviewed
first, then the correspondence between the multipole-Taylor parameters
and the surface mass density multipoles will be given.
When $\sigma(\rvec)$ is expanded in a multipole series as in
equation~\ref{eq:multipole-series}, it can be shown that:
\begin{eqnarray}
\label{eq:multipole-comps-of-sigma-by-int-over-sigma}
\label{eq:multipole-comps-of-sigma-by-int-over-sigma-monop}
\sigma_0(r) &=& \frac{1}{2\pi} \intztp d\theta \sigma(r,\theta)
         \\
\label{eq:multipole-comps-of-sigma-by-int-over-sigma-multip}
\sigvec_m(r) &=& \frac{1}{\pi} \intztp d\theta \sigma(r,\theta) \ehatmlongparen 
.
\end{eqnarray}
These expressions make the physical meaning of the multipoles clear.
The monopole moment $\sigma_0(r)$ is the average surface mass density in
an infinitesimally narrow annulus of radius $r$, or, equivalently,
the angularly-averaged surface mass density.
The multipoles $\sigvec_{m}(r)$ describe
the distribution of matter around that annulus.
In particular, the dipole moment
$\sigvec_{1}(r)$ 
points to the center of mass $\xvec_{c}$ of the annulus, viz.,
\begin{equation}
\frac{\xvec_{c}}{r} = \frac{1}{2} \frac{\sigvec_{1}(r)}
{\sigma_0(r)}.
\end{equation}
Likewise, the quadrupole moment $\sigvec_{2}(r)$ arises from an 
elongated mass distribution. The $\sigvec_{3}(r)$ term
arises from any triangularity in the mass distribution.
Any quadrangularity, e.g. boxiness or diskiness, will give rise to a
nonzero $\sigvec_{4}(r)$ term.
The multipole
$\sigvec_{m}(r)$ can also be expressed as a magnitude $|\sigvec_{m}(r)|$ and an angle
$\massDensAngM(r)$,
\begin{equation}
\sigvec_m(r) = |\sigvec_m(r)| 
( \xhat \cos m \massDensAngM(r) + \yhat \sin m \massDensAngM(r) )
,
\end{equation}
rather than by its $\xhat$ and $\yhat$ components.
Twisted isodensity contours cause
a change in the angle $\massDensAngM(r)$ with radius.
In particular,
a relative excess of mass at radius $r$
in any of the $m$ directions $\massDensAngM(r) + 2\pi n/m$
(where $n\leq m$ is a positive integer)
will contribute to the multipole moment $\sigvec_{m}(r)$.
It follows from the positivity of the surface mass
density that $|\sigvec_m(r)| \leq 2 \sigma_0(r)$.
The multipole moment attains its maximum amplitude only if all the
mass in the annulus at radius $r$ is clumped at the equally spaced
angles $\massDensAngM(r) + 2 \pi n / m$,
though it need not be evenly divided between these angles.
In this manner, knowledge of the mass density's multipole moments gives
a direct picture of the location of the mass.

We now relate the expansion parameters of the
lens potential to the multipole moments of the
surface mass density, by expanding
equations~\ref{eq:poisson-monopole}
and~\ref{eq:poisson-multipole} in a Taylor series about $r = b$. For the
monopole, 
\begin{equation}
\ftWith{t} \equiv \left. \tthderivOneDOverBSqr \Phi_0(r) \right|_{r=\bE}
,
\end{equation}
of which the first four terms are:
\begin{eqnarray}
\ftWith{0} \equiv
\zerothderivOneDOverBSqr \Phi_0(\bE) 
        &=& {\rm const}
        \\
\label{eq:define-einstein-radius}
\ftWith{1} \equiv
\left. \firstderivOneDOverBSqr \Phi_0(r) \right|_{r=\bE}
        &=& 1 
        \\
\label{eq:mass-sheet-term}
\label{eq:coefficient-m0-t2-related-to-surface-mass-density-at-ring}
\ftWith{2} \equiv
        \left. \secondderivOneDOverBSqr \Phi_0(r) \right|_{r=\bE}
        &=& - 1 + \frac{2 \sigma_0(\bE)}{\sigcrit}
        \\
\label{eq:coeff-f3-meaning}
\ftWith{3} \equiv
        \left. \thirdderivOneDOverBSqr \Phi_0(r) \right|_{r=\bE}
        &=& 2 - \frac{2 \sigma_0(\bE)}{\sigcrit} 
            + 2 \left. {{\bE \frac{d}{dr}}} 
                   \left( \frac{\sigma_0(r)}{\sigcrit} \right) \right|_{r=\bE}
\end{eqnarray}

As mentioned previously, the constant term $f_0$ 
can be ignored because a constant offset in the potential does not
affect light deflection, nor does it affect the differential time
delay between two images of the same source.  (Although $f_0$ does affect
the propagation time of the light from source to observer,
only the relative or differential delay
between two different images can be measured.)
The linear term, $f_1$, sets the ring radius $b$ and 
does not have any other adjustable coefficient.
In fact, equation~\ref{eq:define-einstein-radius}
defines the Einstein ring radius $b$ not only for this expansion,
but also for a general potential.
It can be shown that the average surface density
within the radius $b$ so defined
is the critical surface density $\sigcrit$ (\cite{schneider}).
The quadratic term $\ftWith{2}$ gives
the angularly-averaged surface mass density at the ring radius. In
the next section we will show that this term cannot
be constrained by lensing phenomena, due to
the mass-sheet degeneracy.
The cubic term $\ftWith{3}$ is therefore the
first constrainable term yielding information about the
radial profile of the mass distribution.

The correspondence between the higher multipoles
of $\Phi(\rvec)$ and $\sigma(\rvec)$
are obtained by differentiation of equation~\ref{eq:poisson-multipole}.
Before presenting the first few terms of this correspondence,
we define two quantities $\Amvec$ and
$\Bmvec$, which are attributable to the \mth\ multipoles
of the mass that is, respectively, exterior and interior
to the Einstein ring radius:
\begin{eqnarray}
\Amvec & = & \frac{b^{m-2}}{m} \int_{b}^{\infty} dr \, r^{1-m}
                  \frac{\sigvec_{m}(r)}{\sigcrit}
              \\
\label{eq:define-b}
\Bmvec & = & \frac{b^{-m-2}}{m} \int_{0}^{b} dr \, r^{1+m}
                  \frac{\sigvec_{m}(r)}{\sigcrit} 
.
\end{eqnarray}
With these definitions, the correspondence between the higher multipoles
may be written compactly.
In general, for $m\geq1$,
\begin{equation}
\Fmtvec \equiv - \left. \tthderivOneDOverBSqr \Phivec_m(r) \right|_{r=\bE}
.
\end{equation}
Explicitly, the first four terms of the Taylor expansion are:
\begin{eqnarray}
\label{eq:phi-m0}
\Msummvec \equiv \Fvec_{m0} \equiv - \zerothderivOneDOverBSqr \Phivec_m(\bE) 
        &=&  \Amvec + \Bmvec 
        \\
\label{eq:phi-m1}
m \Mdiffmvec \equiv \Fvec_{m1} \equiv 
		- \left.\firstderivOneDOverBSqr \Phivec_m(r)\right|_{r=\bE}
        &=&  m (\Amvec - \Bmvec ) 
        \\
\label{eq:phi-m2}
\label{eq:Fmt2-dependence-on-Am-and-Bm}
\Fvec_{m2}
        \equiv - \left. \secondderivOneDOverBSqr \Phivec_m(r) \right|_{r=\bE}
        &=&       m(m-1) \Amvec + m(m+1) \Bmvec 
                - \frac{2 \sigvec_m(\bE)}{\sigcrit}
        \\
\label{eq:phi-m3}
\label{eq:Fmt3-dependence-on-Am-and-Bm}
\Fvec_{m3}
        \equiv - \left. \thirdderivOneDOverBSqr \Phivec_m(r) \right|_{r=\bE}
        &=&  m(m-1)(m-2) \Amvec - m(m+1)(m+2) \Bmvec 
 \nonumber \\ & & \;\;\;\;\;\;
                        +  \frac{2 \sigvec_m(\bE)}{\sigcrit}
                        - 2 \left. \bE \frac{d}{dr}
                                \frac{\sigvec_m(r)}{\sigcrit} \right|_{r=\bE}
.
\end{eqnarray}
Since the parameter of the 
$t=0$ term is the sum of the exterior and interior
multipoles $\Amvec$ and $\Bmvec$, we label
it $\Msummvec$ rather than $\Fvec_{m0}$.
Likewise, the parameter of the $t=1$ term,
labeled $\Mdiffmvec$, is the
difference between the exterior and interior multipoles.
Higher-order terms ($\Fmtvec$ for $t\geq2$) depend on $\Amvec$ for 
$t\leq m$, and on $\Bmvec$ for all $t$, as well as on the 
behavior of $\sigvec_m(r)$ near $r=b$.
In the next section it will be shown how to modify this
expansion in order to separate explicitly the 
effects of
$\Amvec$ and $\Bmvec$
from the contributions that depend on 
$\sigvec_m(r)$ in the vicinity of  $r=b$.

\subsection{Separation of internal and external contributions}
\label{sec:separation-of-contributions}

The physical meaning of our parameterization becomes clearer if
$\Amvec$ and $\Bmvec$ are used directly as parameters, instead of
$\Msummvec$ and $\Mdiffmvec$.
Furthermore, the sum over $t$ of all the terms
involving $\Amvec$ and $\Bmvec$
can be calculated exactly, leaving only the
effect of the mass near the ring radius left in the Taylor series. 
When this is done,
the resulting ``modified'' multipole-Taylor expansion is:
\begin{eqnarray}
\label{eq:modified-multipole-taylor-series}
\frac{1}{\bE^2}\Phi(r, \theta) & = &
            {\rm const} 
                + \rho 
                + \frac{1}{2} \rho^2 \ftWith{2}
                + \Sumfrom{t=3} \frac{1}{t!} \rho^t \ft
\\ & &
                + \Sumfrom{m=1}
                \left\{ 
                        -\left(\frac{r}{b}\right)^m \Amvec
                        -\left(\frac{r}{b}\right)^{-m} \Bmvec
                        + \Sumfrom{t=2} \frac{1}{t!} \rho^t \Gmtvec
                \right\} \cdot \ehatmlongparen
\nonumber 
\end{eqnarray}
In this expression, the terms $\Gmtvec$ are the portions of $\Fmtvec$
that are left when the contributions proportional to $\Amvec$ and
$\Bmvec$ are subtracted out. The generic form for $\Gmtvec$ is
($m\geq1$, $t\geq2$):
\begin{equation}
\Gmtvec =
        \left. 
                \tthderivOneDOverBSqr \Phimnearring(r) 
        \right|_{r=\bE}
\end{equation}
where
\begin{equation}
\Phimnearring(r) 
 =      - \frac{1}{m} \left(\frac{r}{\bE}\right)^{m}
                \int_r^\bE \left(\frac{r'}{\bE}\right)^{1-m}
                                \frac{dr'}{\bE} \,\,
                                \frac{\sigvec_m(r)}{\sigcrit}
        - \frac{1}{m} \left(\frac{r}{\bE}\right)^{-m}
                \int_\bE^r \left(\frac{r'}{\bE}\right)^{1+m}
                                \frac{dr'}{\bE} \,\,
                                \frac{\sigvec_m(r)}{\sigcrit}
\end{equation}
Explicitly, the first four terms in the Taylor expansion are:
\begin{eqnarray}
\Gvec_{m0}
        &=& 0
        \\
\Gvec_{m1}
        &=& 0
        \\
\label{eq:Gmt2-meaning}
\Gvec_{m2}
        &=& 2\frac{\sigvec_m(\bE)}{\sigcrit}
        \\
\Gvec_{m3}
        &=&             2 \left. \bE \frac{d}{dr}
                                \frac{\sigvec_m(r)}{\sigcrit} \right|_{r=\bE}
                        - 2 \frac{\sigvec_m(\bE)}{\sigcrit}
\end{eqnarray}
We see that these parameters depend only on the distribution of the
deflector mass near the ring radius.
In particular, $\Gvec_{m2}$ gives the \mth\ multipole moment of the 
matter located at the ring radius itself.

\subsection{Degeneracies of the lens equation}
\label{sec:degeneracies}

Certain parameters in the multipole-Taylor expansion cannot be constrained
by observations of gravitational lensing, due to the so-called
``degeneracies'' of the lens equation. A degeneracy, as explained
by Gorenstein, Falco \& Shapiro (1988),
is a family of distinct lens potentials and source
distributions that all produce the same image configuration.
Thus the observation of a given image configuration
does not permit the actual lens potential and source distribution 
to be deduced, but rather only the degenerate family to which they belong.
Breaking the degeneracy requires either additional assumptions
or direct observations of the deflector itself
(such as velocity dispersion measurements; see \cite{dispersion}).

The unconstrainable parameters must be fixed during the minimization of the
error function $\chi^2$, or else convergence is impossible. This section
will identify those parameters and the fixed values we have chosen for them.
The first degeneracy involves the choice of origin.
From equations~\ref{eq:define-b}
and~\ref{eq:multipole-comps-of-sigma-by-int-over-sigma-multip}
follows, for $m=1$,
\begin{equation}
\BmvecWith{1} = \frac{1}{b^3} \int_0^b dr \, r^2
\frac{\sigvec_{1}(r)}{\sigcrit} =
\frac{\xvec_{\rm com}}{b},
\end{equation}
where $\xvec_{\rm com}$ is the center of mass of all the
mass interior to $b$. This parameter
changes with the choice of origin, but obviously the arbitrary choice of
origin cannot affect the image configuration.
Therefore one may allow either $\BmvecWith{1}$ or
the location of the origin to vary, but not both simultaneously.
We decided to impose $\BmvecWith{1} = 0$ and allow the location of the
origin to vary during the minimization procedure. This choice
forces the center of the interior ($r<b$) mass
to be at the origin, thereby simplifying the physical interpretation
of the higher multipoles.

The second degeneracy is the ``prismatic'' degeneracy: any term of the form
$\cvec\cdot\rvec$ in the potential, where $\cvec$ is a constant vector,
is unconstrainable (\cite{dispersion}; \cite{degen}).
The contribution of the exterior dipole $\AmvecWith{1}$ to the potential
(equation~\ref{eq:multipole-taylor-series} 
or equation~\ref{eq:modified-multipole-taylor-series}) is the additive term
$-b\AmvecWith{1}\cdot\rvec$,
which is precisely a prismatic-degenerate term. Consequently, 
we may arbitrarily set $\AmvecWith{1}$ to zero during the model-fitting
procedure.
(If the expansion in equation~\ref{eq:multipole-taylor-series}
is used, rather than the ``modified'' expansion with internal and external
contributions separated out, then the 
center-of-mass degeneracy and the prismatic
degeneracy can be taken into account by
omitting the $\MsummvecWith{1}$ and $\MdiffmvecWith{1}$ terms, 
which is equivalent to
setting both $\AmvecWith{1}$ and $\BmvecWith{1}$ to zero.)

The last and most problematic
degeneracy is the ``mass-sheet degeneracy'' (\cite{dispersion}; \cite{degen}).
Two potentials $\Phi(\rvec)$ and $\Phi'(\rvec)$ cannot be distinguished by
the observation of lensed images if they are related by the transformation:
\begin{equation}
\label{eq:mass-sheet-transformation-of-potential}
\Phi'(\rvec) = (1-\kappa)\Phi(\rvec) + \frac{\kappa r^2}{2}.
\end{equation}
This is equivalent to reducing the surface mass density by a factor $(1-\kappa)$
and adding a sheet of uniform mass density $\kappa\sigcrit$:
\begin{equation}
\sigma'(\rvec) = (1 - \kappa) \, \sigma(\rvec) + \kappa \sigcrit .
\end{equation}
The corresponding relations between the multipole-Taylor parameters
of the original and transformed potential are:
\begin{eqnarray}
	&& f'_0 = \rescale f_0 + \frac{1}{2} \kappa 
		= {\rm const}'\\
	&& f'_1 = f_1 = 1 \\
	&& f'_2 = \rescale \ftWith{2} + \kappa \\
t\geq3: && f'_t = \rescale \ftWith{t} \\
	&& \\
	\label{eq:mass-sheet-degeneracy-transformation}
	&& \Amvec' = \rescale \Amvec \\
	&& \Bmvec' = \rescale \Bmvec \\
	&& \Gmtvec' = \rescale \Gmtvec 
\end{eqnarray}
The constant offset has been altered, but this is immaterial since
adding a constant to the potential has no effect on light
deflection or differential time delays.  The transformation
must leave the ring radius unchanged, so the coefficient of
$\rho$ in the monopole part of the potential is unchanged.  The
quadratic monopole term, $f_2$, which depends on the surface mass density at
the ring radius (Eq.~\ref{eq:mass-sheet-term})
has been changed, because the surface mass density at the ring radius
has been changed.  All the coefficients of the other terms in the
model have been scaled by a factor $1 - \kappa$.  These simultaneous
adjustments of all the $\ftWith{t}$ ($t\geq2$),
$\Amvec$, $\Bmvec$, and $\Gmtvec$ parameters leave the model
predictions unchanged.
Therefore one of these parameters should be fixed in some manner
during the model-fitting procedure
to permit convergence of the minimization algorithm.

We note that the transformation of 
equation~\ref{eq:mass-sheet-transformation-of-potential}
does affect the time delays,
\begin{equation}
	\Delta t' = \rescale \Delta t
.
\end{equation}
This however does not permit observations of the time delays to be
used to determine the scaling factor $(1-\kappa)$, since a rescaling
of the time delay will be confused with a rescaling of the Hubble
parameter.
We discuss this issue further in section~\ref{sec:implications-for-time-delays}.

To break the degeneracy, and permit convergence of the minimization
algorithm, 
we may arbitrarily select a value of $f_2$;
by equation~\ref{eq:mass-sheet-term},
this is equivalent to an arbitrary choice of $\sigma_0(b)$.
We find it convenient to fix $f_2' = 0$, which happens to be correct for an
isothermal sphere, for which $\sigma_0(b) = \sigcrit/2$.
If this is not true of the 
actual 
mass distribution, then the
fitted (primed) 
amplitudes of all the 
$m\neq0$ multipoles and the $t\geq3$ monopole parameters 
will have been scaled by the same factor $1-\kappa$
relative to the true (unprimed) amplitudes,
where
\begin{equation}
1-\kappa = \frac{1}{2\left(1-\frac{\sigma_0(b)}{\sigcrit}\right)}
\end{equation}
These scaling factors are tabulated in 
Table~\ref{tbl:comparison-to-other-models} for several
different choices of the actual monopole potential.
For example, suppose the choice $f_2' = 0$
is made, but the true potential is actually a point mass rather
than an isothermal sphere. Then, in reality, $\Phi_{0}(\rvec) = b^{2}\ln r$,
and $\sigma_0(b) = 0$. All of the best-fit parameters 
($f_t'$, $\Amvec'$, etc.)
and the model time delays ($\Delta t'$)
are smaller than those 
($f_t$, $\Amvec$, $\Delta t$, etc.)
of the actual mass distribution,
having  been scaled by the same factor $(1-\kappa) = 1/2$.
Note that this rescaling only affects the magnitudes of the multipoles;
it does not affect their angles.
It also does not affect the center-of-mass parameters $g_x$ and $g_y$,
or the ring radius $b$.

Once we have set $f_2' = 0$, the next term in the monopole expansion,
$\ftWith{3}'$, is the lowest-order monopole parameter available to
give information on the radial distribution of mass.
After rescaling, its amplitude is related to the true surface mass density by
\begin{equation}
\label{eq:coeff-f3prime-meaning}
\ftWith{3}' = 1 - \left. \firstderivOneD \ln \left( 
                        1 -  \frac{\sigma_0(r)}{\sigcrit}
                                        \right) \right|_{r=\bE}
\end{equation}
in contrast to equation~\ref{eq:coeff-f3-meaning}.
It depends on the fall-off of the surface mass density 
near the Einstein ring radius, and is
most sensitive to that falloff when $\sigma_0(\bE)$ is close
to $\sigcrit$.
For surface mass densities
that do not increase with radius, $\ftWith{3}' < 1$.
The values of $\ftWith{3}'$ for various monopole
potentials are given in Table~\ref{tbl:comparison-to-other-models}.
The lowest value for
$\ftWith{3}'$ listed in the table is $-1$, for a
mass sheet, but $\ftWith{3}'$ can be even more
negative for potentials with a core radius, or
if $\sigma_0(r)$ drops abruptly near the ring radius.

The mass-sheet degeneracy also affects the interpretation of the
amplitude of the quadratic multipole parameter $\Gvec_{m2}'$, 
whose fitted value is
related to the true surface mass density by
\begin{equation}
\Gvec_{m2}' = \frac{\sigvec_m(\bE)}{\sigcrit - \sigma_0(\bE)},
\end{equation}
in contrast to equation~\ref{eq:Gmt2-meaning}. 
Non-negativity of the surface mass density
limits the amplitude of $\Gvec_{m2}'$:
\begin{equation}
\label{eq:limit-on-amp-of-Gmt2}
|\Gvec_{m2}'| \leq \frac{2\sigma_0(\bE)}{\sigcrit-\sigma_0(\bE)}
.
\end{equation}
This limit ranges from zero, for $\sigma_0(\bE) = 0$,
to $+\infty$, for $\sigma_0(\bE) = \sigcrit$.
For a mass distribution with a singular isothermal sphere monopole,
the limit is $|\Gvec_{m2}'| \leq 2$.,
For mass distributions more centrally concentrated than a
singular isothermal sphere, the limit on
$|\Gvec_{m2}'|$ would be lower.
The limit is only attained if all the mass at the ring radius is clumped
into point-like perturbers in any of the $m$ allowed directions.

In summary, after expanding the lens potential in a multipole-Taylor
series, then explicitly separating the contributions from the internal
and external multipoles, and then fixing several parameters to zero
(as described above) because of the degeneracies in the lens equation,
the final parameterized form of the potential is:
\begin{eqnarray}
\label{eq:final-form-of-potential}
\frac{1}{\bE^2}\Phi'(r, \theta)
        &=&       \rho 
                + \Sumfrom{t=3} \frac{1}{t!} \rho^t \ft'
\nonumber \\ & &
                - \Sumfrom{m=2}
                \left\{ 
                        (1+\rho)^m \Amvec'
                        + (1+\rho)^{-m} \Bmvec'
                \right\} \cdot \ehatmlongparen
\nonumber \\ & &
                + \Sumfrom{m=1}
                        \Sumfrom{t=2} \frac{1}{t!} \rho^t \Gmtvec'
                        \cdot \ehatmlongparen
. \nonumber \\ & &
\end{eqnarray}
Here and elsewhere, whenever the distinction is important,
we have used primes to identify fitted model
parameters, reserving the non-primed symbols
for the parameters that describe the actual gravitational potential.
These two sets of parameters differ because of the arbitrary
degeneracy-breaking choices described in this section.
Equation~\ref{eq:final-form-of-potential}
is the parameterization we used to compute
best-fit models for MG~J0414+0534 (although,
in a few cases, we tried using 
the parameter $\Msummvec'$ from the original multipole-Taylor series,
equation~\ref{eq:multipole-taylor-series}, in place of $\Amvec'$ and $\Bmvec'$).
A summary of the physical significance of each term in this expansion
is given in Table~\ref{tbl:physical-significance}.
Since the analytic forms of the first and second derivatives of the
potential are used during the model-fitting procedure, these derivatives 
are presented in Tables~\ref{tbl:first-derivatives}
and~\ref{tbl:second-derivatives}.

\subsection{Comparison to other parameterizations}

Any parameterized form of the lens potential $\Phi(\rvec)$ can be
compared to ours by expanding it in a modified multipole-Taylor series.
In this section we 
carry out this procedure for a few forms for
the lens potential that are widely used.

Often the lens potential is taken to be a combination of a
monopole term and a quadrupole term, just as Kochanek (1991) did.
The particular form of each of these is
usually prescribed by either physical preconceptions or ease of computation.
In Table~\ref{tbl:comparison-to-other-models}
we compare five possible monopole potentials. These all give rise to
the same linear term which sets the Einstein ring radius.
The first significant term that distinguishes
them is therefore $f'_3$, since $f_2$ is not constrainable due to
the mass-sheet degeneracy.
If the data are not sufficient to distinguish among the various possibilities
for $f'_3$, then there is little point in using more complicated forms
for the monopole term,
such as de Vaucouleurs radial profiles (\cite{ellithorpe}).

There are three different quadrupole terms that are commonly used to accompany
the monopole term.
The first is an external quadrupole, of the form:
\begin{equation}
\Phi_{Q}(r,\theta) = \frac{\gamma}{2} r^2 \cos 2(\theta-\theta_{\gamma})
\end{equation}
This term is sometimes referred to as an ``external shear.'' In our scheme,
this is equivalent to choosing a non-zero value of $\AmvecWith{2}$, with
magnitude $|\AmvecWith{2}| = \gamma/2$ and angle 
$\psi_{A_2} = \pi/2 + \theta_{\gamma}$. (Our notation has $\psi_{A_2}$ and
$\psi_{A_2} + \pi$ as the angles to the mass excess, whereas 
$\theta_{\gamma}$ and $\theta_{\gamma} + \pi$ are the angles to the 
mass deficit.)
The second is an internal quadrupole term, parameterized as:
\begin{equation}
\Phi_{Q}(r,\theta) = -\frac{\gamma}{2} \frac{b^4}{r^2} \cos 2(\theta-\theta_{\gamma}),
\end{equation}
which is equivalent, in our scheme, to choosing a non-zero $\BmvecWith{2}$ with
magnitude $|\BmvecWith{2}| = \gamma/2$ and angle $\psi_{B_2} = \theta_{\gamma}$. 
Finally, there is the case of a ``mixed'' quadrupole,
\begin{equation}
\Phi_{Q}(r,\theta) = -\frac{b\gamma}{2} r \cos 2(\theta-\theta_{\gamma}),
\end{equation}
which is obtained by truncating the potential of a singular isothermal 
elliptical potential at
the quadrupole term. To order $\rho$, this is equivalent to the choice
$|\AmvecWith{2}| = 3\gamma/8$, 
$|\BmvecWith{2}| = \gamma/8$, and 
$\psi_{A_2} = \psi_{B_2} = \theta_{\gamma}$.
It is worth noting here that 
$\AmvecWith{2} + \BmvecWith{2}$ causes tangential image displacements for
images near the ring radius, whereas $\AmvecWith{2} - \BmvecWith{2}$ causes
radial image displacements. 
The balance between the internal and external portions of the quadrupole 
therefore determines
the radial displacement that accompanies the
tangential displacements caused by the quadrupole moment
of the mass distribution.
The choice of a ``mixed'' quadrupole is
essentially a particular choice for this ratio.

\subsection{Effects of external perturbing masses}

If there is a deflecting object along the line of sight to the
source besides the primary lensing galaxy, this secondary deflector
can be modeled in one of two ways. It could be treated separately
from the primary deflector, with extra parameters for its location
and mass distribution.
Alternatively, the parameterization of
equation~\ref{eq:final-form-of-potential}
could be used alone, so that the influence of the secondary perturber would
be reflected in the values of the multipole moments.
Since the location (or even the presence) of a perturbing object is
not known {\em a priori}, it is useful to compute the effect
on the parameters of the multipole-Taylor expansion that would be
caused by a perturbing object far removed from the primary deflector.

A perturbing point mass located further from the origin than any of the lens
images contributes to the external multipole moments to 
all orders $m$, with amplitudes
\begin{equation}
\label{eq:external-multipole-moment-of-PM-perturber}
|\Amvec|
= \frac{1}{m} \left(\frac{\bE}{R}\right)^m \left(\frac{b_P}{\bE}\right)^2 
,
\end{equation}
where $b$ is the ring radius of the principal deflector, $b_P$ is the
ring radius of the perturber, and $R$ is the distance to the perturber.
This can be derived by expanding its delta-function surface-mass distribution
as a multipole-Taylor series.
The $m=1$ term has no effect, due to the prismatic degeneracy, so the
dominant term is the quadrupole moment, which is often called the external shear.
For small $\bE/R$, it should be adequate to represent the
effect of the perturber by the first few multipoles.
Conversely, the ratios of the fitted amplitudes $\AmAmp$
(if they can be attributed solely to a perturbing influence)
permit the distance to the perturber, and its strength, to be deduced.

How do radial and tangential extent in an external perturber affect
the external multipole amplitudes?
Consider a perturber with its center of
mass located at a distance $R$ from the center of the principal
deflector, with surface mass density uniform over a region
of radial extent $\Delta R$ and tangential extent $R \Delta \theta$.
Such a perturber contributes to the external multipole moments of the
potential to all orders $m$ in the multipole expansion, with
amplitudes
\begin{eqnarray}
\label{eq:external-multipole-moment-of-slightly-extended-perturber}
|\Amvec|
= \frac{1}{m} \left(\frac{\bE}{R}\right)^m \left(\frac{b_P}{\bE}\right)^2 
\left\{ \rule{0cm}{3.5ex}\right. 
1 
  & \!\! + \!\! & 
   \frac{m(m+1)}{24}\left( \frac{\Delta R}{R} \right)^2
  - \frac{m(m+1)}{24}\left(     \Delta\theta   \right)^2
\nonumber \\ 
  & \!\! + \!\! &   
          \orderof\left( \frac{\Delta R}{R} \right)^4
        + \orderof\left(     \Delta\theta   \right)^4
        + \orderof\left(     \Delta\theta   \right)^2
            \orderof\left( \frac{\Delta R}{R} \right)^2
\left.\rule{0cm}{3.5ex} \right\}
,
\nonumber \\
\end{eqnarray}
The net effect, for perturbers with similar radial
and tangential extents ($\Delta R = R \Delta \theta$) is that there
is no effect through $\orderof\left(\frac{\Delta R}{R}\right)^3$.
Therefore, using the formulae for a point mass perturbation
(eq.~\ref{eq:external-multipole-moment-of-PM-perturber})
should cause no problem for a perturber that is located far 
away as compared to its extent.
Unfortunately, for a nearby extended perturber, the interpretation of
the fitted external multipole amplitudes is less simple.

\section{Application to MG~J0414+0534}
\label{sec:application}

A model of a gravitational lens consists of two parts: a model of the surface brightness
of the source, as it would appear in the absence of lensing,
and a model of the lens potential.
An ideal model reproduces the observed image of the system, pixel for pixel,
by mapping the source surface brightness through the lens potential and then convolving
with the detector response. This is the aim of modeling techniques such as LensClean
(\cite{lclean}; \cite{vlclean}), which use the information from every pixel to
constrain the model.

It is computationally faster to employ a much smaller number of constraints that
capture the most important features of the observed image.
Multiple images of nearly-pointlike sources,
such as the 16 components seen in MG~J0414+0534,
are succinctly described by the locations and flux densities of the
pointlike components. With this simplification, called ``point modeling,'' a much wider
range of models can be explored.

We chose to use only the positions of the 16 components of MG~J0414+0534, and not
their flux densities, as model constraints. There were two reasons for this.  One, the
relative uncertainties in the fluxes are
much larger than the relative uncertainties in the centroid positions.
Therefore the inclusion of flux information would not contribute much to
the error statistic $\chi^2$.
Two, there are significant discrepancies between the radio measurements and
optical measurements of the flux density ratios of components in MG~J0414+0534.
This discrepancy has been variously attributed to microlensing
(\cite{wittmao}), variable extinction (\cite{dust}), and/or substructure in the lensing galaxy (\cite{substructure}).
Some of these effects may not
be significant for radio observations, but we chose not to use flux
density information at all.

\subsection{Model-fitting algorithm}
\label{sec:algorithm}

Our constraints consist of the observed positions $\rvec_{k\alpha}$, where the
index $\alpha$ runs over the 4 different images 
(A1, A2, B, and C)
and the index $k$ runs over the 4 components 
(p, q, r, and s)
of each image. The model consists of presumed source positions $\svec_{k}$
for each component, along with the mapping $\rvec_{\alpha}(\svec_{k})$
provided by the lens equations 
(eq.~\ref{eq:lens-equation})
using our expansion of the potential 
(eq.~\ref{eq:final-form-of-potential}) truncated to a prescribed order.

Assuming that the observational errors are Gaussian, the maximum-likelihood values
of the parameters can be determined by minimizing the familiar chi-squared statistic,
\begin{equation}
\chi^2 = \sum_{k=1}^4 \sum_{\alpha=1}^4 (\rvec_{\alpha}(\svec_{k}) - \rvec_{k\alpha})
\cdot {\bf S}^{-1}_{k\alpha} \cdot
(\rvec_{\alpha}(\svec_{k}) - \rvec_{k\alpha}),
\end{equation}
where ${\bf S}_{k\alpha}$ is the error covariance matrix. Because the lens mapping is much easier to
apply in the direction $\rvec \rightarrow \svec$ than the reverse direction, it is
of great computational advantage
to compute an approximate value of $\chi^2$ by evaluating the model errors
in the source plane rather than the image plane (\cite{kayser}).
By Taylor expansion, the displacement between the model
source position and the actual source position is 
\begin{equation}
\svec_k - \svec(\rvec_{k\alpha}) = 
      {{\bf M}^{-1}}(\rvec_{k\alpha})
        (\rvec_\alpha(\svec_k) - \rvec_{k\alpha})
        + 
        \left( (\rvec_\alpha(\svec_k) 
              - \rvec_{k\alpha}) \cdot \Del_{\rvec}
                       {{\bf M}^{-1}}(\rvec_{k\alpha}) \right)
                                                (\rvec_\alpha(\svec_k) - \rvec_{k\alpha})
        + \ldots ,
\end{equation}
where ${\bf M^{-1}}(\rvec_{k\alpha})$ is the inverse magnification matrix,
\begin{equation}
\label{eq:inv-mag-matrix}
{M^{-1}}_{ij}(\rvec) = \delta_{ij} - \frac{\partial}{\partial
r_i} \frac{\partial}{\partial r_j} \Phi(\rvec)
.  
\end{equation}
For good models,
the difference between the modeled and observed image-plane positions,
$\rvec_{\alpha}(\svec_{k}) - \rvec_{k\alpha}$, is small, so the
higher order terms in the expansion may be neglected.
Equivalently, the change in magnification
between the observed image location and the model-predicted image location
is assumed to be negligible.
The resulting ``source-plane'' approximation for $\chi^2$ is:
\label{sec:source-plane-chi}
\begin{equation}
\chi^2 \approx \chi^2_{s}
= \sum_{k=1}^4 \sum_{\alpha=1}^4 (\svec_{k} - \svec(\rvec_{k\alpha}))
{\bf M}(\rvec_{k\alpha})
\cdot {\bf S}^{-1}_{k\alpha} \cdot
{\bf M}(\rvec_{k\alpha})
(\svec_{k} - \svec(\rvec_{k\alpha}))
\end{equation}

This approximation is useful because
it is valid near the chi-squared
minimum, and we are unconcerned
with the behavior of the function far from the minimum.
The global minimum of both $\chi^2$
and $\chi^2_{s}$ is zero,
which occurs when the model exactly reproduces the observation.  
We expect that even with noise and measurement
error, $\chi_{s}^2$
has a global minimum corresponding to that
of the true chi-squared, $\chi^2$, 
and that no lower minimum is introduced by this
approximation.
This approximation fails for very poor deflector
models that do not reproduce
the image locations at all --- but this is of little
concern, since the high $\chi^2_{s}$ would cause the model to
be rejected anyways.
The approximation can also fail
if the error in the observed image locations is large
enough to encompass a region in which the magnification matrix
varies significantly.
Because of this danger, after the 
minimum of $\chi^2_s$ was found for each model,
we computed the true $\chi^2$ to check
the source-plane approximation. 
For models that adequately satisfied the observational constraints
(as described below), $\chi^2_{s}$ typically differed from
$\chi^2$ by only $0.2-0.6\%$, and at most $1.5\%$.
Since this is much less that the $\chi^2$ increment
used to find confidence limits on model parameters,
the source-plane approximation introduced no appreciable error.

Since the first and second derivatives of our parameterized
potential are available in analytic form, the magnification
matrix and thus $\chi^2_{s}$ are easy to compute.
These derivatives are listed in Tables~\ref{tbl:first-derivatives} 
and~\ref{tbl:second-derivatives}.
Furthermore, since $\chi^2_{s}$ is
quadratic in the source positions $\svec_{k}$,
the optimal source positions are easily
computed for given values of the model parameters. Consequently,
the numerical minimization need 
not include a search through the source positions in addition to 
the parameters of the deflector model.
This reduction in the number of dimensions of
the search space permits a vast computational
speed-up.
(If fluxes also are used as model constraints, a
corresponding source-place approximation can still be made. The
resulting $\chi^2_s$ is quadratic in both model source positions and
model fluxes, so the optimal model positions and fluxes can still be
found analytically [\cite{cathythesis}].)

To perform the minimization of $\chi^2_{s}$ we employed a variant
of simulated annealing described in \cite{recipes}. 
Simpler methods, such as a straightforward downhill
walk via the Powell direction set method,
or the downhill simplex method (\cite{recipes}),
encountered difficulties 
with local minima.
The initial value for the center
of mass was set to the galaxy position observed in the HST
images of Falco, Leh\'{a}r \& Shapiro (1997).
The initial value of the
ring radius $b$ was taken from the
previous best-fit models of Ellithorpe (1995),
and the initial values of all $m\geq 1$
multipoles were set to zero.
To explore the region of parameter space near these
physically-motivated starting values,
we used a moderately
low starting temperature equal to 1\% of the initial value of $\chi^2_{s}$.
High starting temperatures allow the possibility of escaping
the physically reasonable portion of parameter space and becoming
trapped in deep and distant local minima.

The higher multipole parameters were
represented by their Cartesian components,
e.g. $\xhat \cdot \AmvecWith{2}'$ and $\yhat \cdot \AmvecWith{2}'$, rather than
by amplitude and angle. The potential depends linearly
on these parameters, and thus the chi-squared depends 
roughly quadratically on them
(far from the minimum, at least), allowing for a more robust
minimization.
However, confidence limits were computed for the
amplitude and angle of
each multipole parameter, rather than its Cartesian components, because
the amplitude-angle representation
is more useful for visualizing the mass distribution,
and because amplitudes are affected by the mass-sheet degeneracy
but angles are not.

For the value of $\chi^2$ to be used to calculate confidence intervals for the model
parameters,
the estimates of the observational errors (as represented in the
covariance matrix ${\bf S}_{k\alpha}$) must be accurate.
Unfortunately, it is difficult to make accurate error estimates of VLBI centroid positions,
because of the complicated and nonlinear process of deconvolution by ``cleaning''
and self-calibration. In addition, if there are magnification gradients across the image,
the image centroid may not be exactly the image of the source centroid, even though the
point-modeling approach assumes so.  
The separation between the image of the source-centroid and the
centroid of the image is approximately the angular size of the
image multiplied by the fractional change in the magnification over the
extent of the image.  (The expression for the discrepancy is given by
Trotter (1998), along with a correction to $\chi^2_{s}$ to account
for it.  We did not use this correction because it requires the
calculation of the third derivatives of the potential and
accurate estimates of the intrinsic source size.)

For these reasons we report the results using three different
methods to estimate the positional error in each component. The first, a
crude upper limit,
is the image size convolved with the VLBA beamwidth, which we call ``fit-size errors.''
The second estimate for the positional error is a lower limit: 
the statistical error in the centroid
position due to thermal noise in the map. We report this as ``statistical error.''
The third estimate is the quadrature sum of the
statistical error and the width of the deconvolved image (the intrinsic
component size).
This ``stat-width error''
makes some allowance for magnification gradients as well as deconvolution error.
We believe the stat-width estimate to be the most accurate of the three estimates,
but since this judgment is not rigorous we report the results for $\chi^2$
using all three estimates.
Confidence limits on each model parameter were determined using the
``stat-width error,'' by stepping the parameter away from 
the $\chi^2_s$ minimum, while minimizing over all other parameters, until
the $\Delta \chi^2$ appropriate for 68.3\% confidence limits was
obtained.

\subsection{Model results}
\label{sec:model-results}

Table~\ref{tbl:results} summarizes the goodness-of-fit for a variety of
model potentials. For each model the number of parameters
and number of degrees of freedom are listed,
along with the minimum $\chi^2$ obtained using each of the three different 
estimates of positional errors. In this section we review these results.

Before using the modified multipole-Taylor expansion, we tested three simpler
parameterizations that have been used in
previous attempts to model MG~J0414+0534 based on
lower-resolution radio and optical data.
These simpler potentials were
a singular isothermal sphere plus external shear (SIS+XS),
a point mass with external
shear (PM+XS), and a singular isothermal elliptical
potential truncated at the
quadrupole moment (SIEP)
(see e.g. \cite{hewitt}, \cite{hst}, \cite{ellithorpe}):
\begin{eqnarray}
\label{eq:SIS-XS-pot}
\Phi_{\rm SIS+XS}(r,\theta) &=&
b r +  \frac{1}{2} \gamma r^2 \cos 2 ( \theta - \theta_\gamma)
\\
\label{eq:PM-XS-pot}
\Phi_{\rm PM+XS}(r,\theta) &=&
b^2 \ln r +  \frac{1}{2} \gamma r^2 \cos 2 ( \theta - \theta_\gamma)
\\
\label{eq:SIEP-pot}
\Phi_{\rm SIEP}(r,\theta) &=&
b r      - \frac{b\gamma}{2}  r \cos 2( \theta - \theta_\gamma)
\end{eqnarray}

The results for these 5-parameter models are shown in the top three 
lines of Table~\ref{tbl:results}.
They are all very poor fits.
Qualitatively they fail to reproduce the detailed VLBI structure of
the four components. Quantitatively, even for the upper-limit
(fit-size) errors, the $\chi^2$ is more that $3 \times 10^3$ standard
deviations away from the value that would be expected for 
observations matching these models. 

The rest of the models in Table~\ref{tbl:results} 
are multipole-Taylor expansions
truncated in various ways. They are labeled with symbols indicating the terms
that are present in the expansion. 
The dominant terms in the multipole-Taylor expansion --- the only terms
that cause shifts in the image positions for images at the ring
radius --- are the first (linear in $\rho$)
monopole term, $b$, and the first
two (constant and linear in $\rho$, or external and internal)
$m\geq2$ multipole terms, $\Amvec$ and $\Bmvec$.
Accordingly, all of the models included the $b$ term, which sets the
Einstein ring radius, as well as the internal quadrupole
($\BmvecWith{2}$) and external shear ($\AmvecWith{2}$) terms which account
for ellipticity of the deflector mass distribution and the dominant
effects of any external perturber.
The model containing these three terms and no
others is given the schematic name
$b+\AmvecWith{2}+\BmvecWith{2}$,
and is the fourth model listed in Table~\ref{tbl:results}.
It differs from the SIS+XS model only by the
addition of the internal quadrupole $\BmvecWith{2}$ term. The addition of this
second quadrupole term causes a vast improvement in the fit;
the minimum $\chi^2$ is lowered by two orders of magnitude,
although the model is still not in formal agreement with the data.

All of the other models are labeled with a ``+'' sign and the terms
they contain in addition
to the three terms $b$, $\AmvecWith{2}$, and $\BmvecWith{2}$.
The next batch of models, as indicated in Table~\ref{tbl:results}, each include
only one term in addition to these three.
The next most significant terms in each of
the $m=0$ through $m=4$ multipoles were tried.
The $m=3$ multipoles were considered because they
are the next terms in which effects of an external perturber
would appear, and would also account for any lopsidedness in the mass
distribution of the lens galaxy.
The $m=4$ multipoles were considered to account for
diskiness or boxiness of the lens galaxy.
Higher multipoles, $m\geq5$, were not tried,
as there was no physical reason to expect them to be significant.
Also tried were the mixed-internal-and-external terms, 
$\MsummvecWith{3}$ and $\MsummvecWith{4}$,
of the original multipole-Taylor expansion,
equation~\ref{eq:multipole-taylor-series}.
Using $\MsummvecWith{m}$ instead 
of $\AmvecWith{m}$ or $\BmvecWith{m}$ strikes
a different balance between the internal and external
contributions to the \mth\ multipole.

In all cases the fits were an improvement over the three-term
model, but the best results were obtained by adding either
an external octupole $\AmvecWith{3}$, or 
a mixed-internal-and-external octupole $\MsummvecWith{3}$.
These two models fit the image positions well enough to satisfy the
fit-size (upper limit) position errors, though not well enough to satisfy the
two tighter error estimates. For both of these models, the angle of
the external quadrupole ($74.4\arcdeg\pm 0.2\arcdeg$ E of N
for $+\AmvecWith{3}'$; $74.1\arcdeg\pm 0.2\arcdeg$ for $+\MsummvecWith{3}'$)
is consistent with the optical isophote angle of
the deflector as observed in the WFPC2 image of
\cite{hst} ($71\arcdeg \pm 5\arcdeg$).
For the $+\AmvecWith{3}$
model, the direction of the
internal quadrupole $\BmvecWith{2}'$ ($75.4\arcdeg\pm 0.5\arcdeg$ E of N)
also agrees with the
WFPC2 optical isophotes.
The best-fit
parameters of this 9-parameter model ($+\AmvecWith{3}$)
are displayed pictorially in Figure~\ref{fig:A3}.
The $+\MsummvecWith{3}$ model has a somewhat
smaller value of $\chi^2$, although in this case
the internal quadrupole
$\BmvecWith{2}'$ ($87.4\arcdeg\pm 0.9\arcdeg$ E of N)
and the isophotes are misaligned by $16\arcdeg$.

To each of these promising models, $+\AmvecWith{3}$ and $+\MsummvecWith{3}$,
was added the next radial term in each $m\leq4$ multipole,
one at a time, as indicated
in the third block of Table~\ref{tbl:results}. 
These models had 10 or 11 parameters,
depending on whether the additional term was monopole or not.
For the sake of comparison, a model with 11
parameters but employing $m=4$ multipoles instead of
$m=3$ multipoles was also tried (and fared very poorly).

The models including the external octupole $\AmvecWith{3}$
outperformed the models including the mixed octupole
$\MsummvecWith{3}$. The model with the lowest $\chi^2$
was
$+\AmvecWith{3}+\AmvecWith{4}$ (external $m=3$ and $4$ multipoles),
followed by the model
$+\AmvecWith{3}+\BmvecWith{3}$ (external and internal $m=3$
multipoles).\footnote{
The model $+\AmvecWith{3}+\Gvec_{22}$ produced the unphysical
value $|\Gvec_{22}'| = 0.5$.
This would require that at least $25\%$ of the
mass in a narrow annulus at $r=b$ be involved
in driving the quadrupole term
--- or more, if mass is not concentrated into points.
(This assumes that the mass
of the deflector is not more extended than a singular isothermal sphere.)
See equation~\ref{eq:limit-on-amp-of-Gmt2}.
To produce sizable image displacements,
$|\Gvec_{22}'|$ must be large 
because it must overcome the small factor $\rho$.
This term is
apparently compensating for the internal and external
quadrupoles, which are displaced relative
to their orientations in the $+\AmvecWith{3}$ model.
}
Both of these models oversatisfy even the stat-width
error estimates (indicating, perhaps, that these error estimates may be
too large), though neither model is formally consistent with the
lower-limit error estimates. The best-fit parameters
for the most successful 11-parameter model, $+\AmvecWith{3}+\AmvecWith{4}$,
are shown pictorially in Figure~\ref{fig:A3_A4}.
The implications of the success of this and other
models will be considered in the next section.

The lowest-order term in the multipole-Taylor expansion
that is both constrainable and sensitive to the radial 
distribution of mass is $f_3'$.
In the next round of modeling we added
the term $f_3'$ to each of the models of the
previous group,
in order to determine whether the radial dependence of the potential
could be usefully constrained.
The results are shown in the last block of Table~\ref{tbl:results}.
The fitted parameter values
have large error ranges 
(e.g. $f_{3}' = 1.13^{+0.39}_{-0.52}$ for the
model $+\AmvecWith{3}+\AmvecWith{4}+\ftWith{3}$)
compared to the range of interesting
values, which extends from $f_{3}' = 0$ (singular isothermal sphere)
to $f_{3}' = 1$ (point mass).
More problematic is that
the value of this parameter depends sensitively on the
presence or absence of the other multipole components in the model,
with values ranging from $-9.7$ to $2.3$ for 
12-parameter models that adequately satisfy the
stat-width error estimates.
It is clear that useful information on the radial profile of
MG~J0414+0534 is unavailable from this data.

\section{Discussion}
\label{sec:discussion}

\subsection{Implications for the mass distribution}
\label{sec:discussion-mass}

\begin{figure}
\plotfiddle{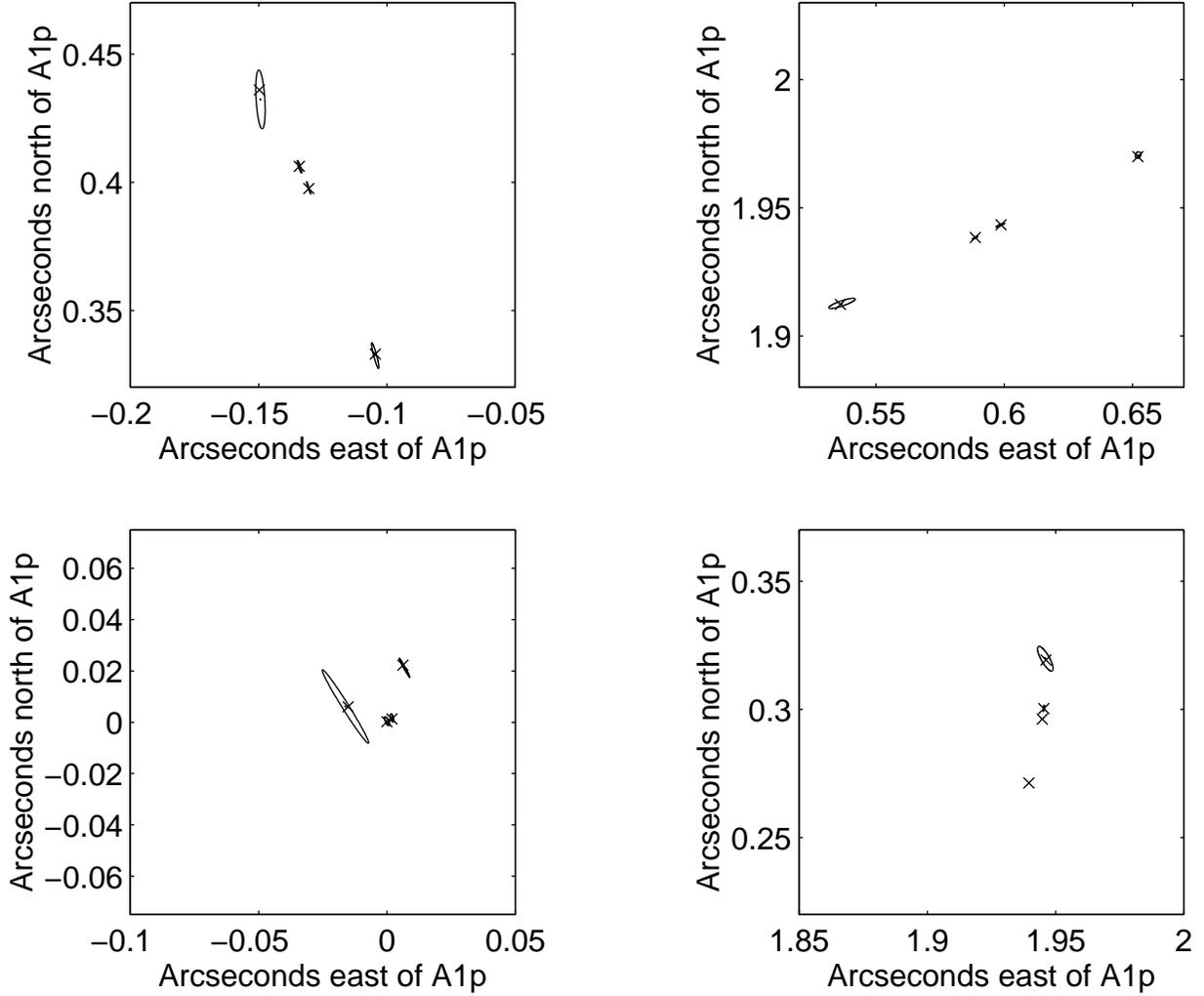}{6in}{0}{100}{100}{-300}{-150}
\caption{
Observed and model positions for the images of MG~J0414+0534.
Clockwise from lower left, the panels show the vicinity of
the images A1, A2, B and C.
The model positions are plotted as small crosses ($\times$).
The error ellipses represent the 90\%-confidence intervals
for the observed image positions. The large
error ellipses were computed using the stat-width error
estimates, whereas the small error ellipses (most of which appear
pointlike on a plot of this scale) were computed using the
statistical error estimates.
The different error estimates are described in \protect\ref{sec:algorithm}.
}
\label{fig:residuals}
\end{figure}

\begin{figure}
\plotfiddle{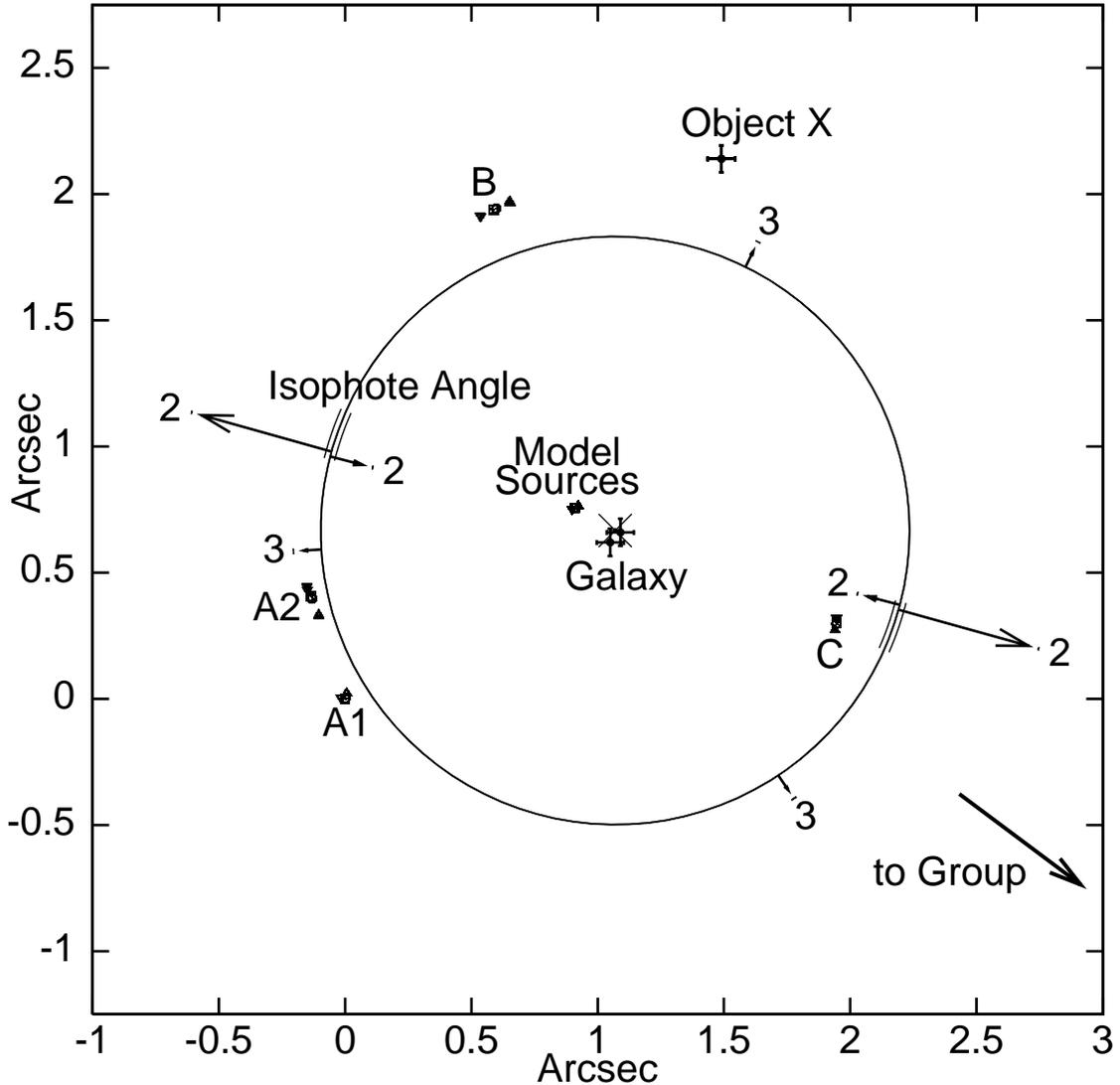}{6in}{0}{200}{200}{-350}{-100}
\caption{
{\small
Model $b + \AmvecWith{2} + \BmvecWith{2} + \AmvecWith{3}$, fitted to MG~J0414+0534.
The model parameters are illustrated graphically.
North is up and east is left.
An $\times$ marks the
model deflector center of mass $(g_x, g_y)$, and a ring is drawn at
the model Einstein ring radius $b$.  The error range on the ring
radius is shown by additional rings drawn at the upper and lower
confidence limits.
(At this plot scale, neither the error rings nor the
error range for the model galaxy position can be resolved.)
Each $m\geq1$ multipole
moment is illustrated by arrows pointing in the possible directions to
mass excesses that would cause such a multipole
moment.  The arrows are labeled with the number $m$ of the multipole
moment, and their lengths are proportional to the amplitudes of the
multipole moments.  Confidence ranges on the angles are indicated by
small arcs at the tip of each arrow.  Also shown are the
$R$ and $I$-band center positions of the deflector galaxy ($+$ marks), 
the position of Object X,
the direction to a nearby group of galaxies (arrow),
and the orientation of the galaxy isophotes
(arcs near the ring radius),
all from Falco, Leh\'{a}r \& Shapiro (1997)
The locations of the observed and
modeled components
are also shown but cannot be distinguished at this plot scale.
See Figure~\protect\ref{fig:residuals} for a closer view.
The model sources are plotted northeast of the galaxy center.
}
}
\label{fig:A3}
\end{figure}

\begin{figure}
\plotfiddle{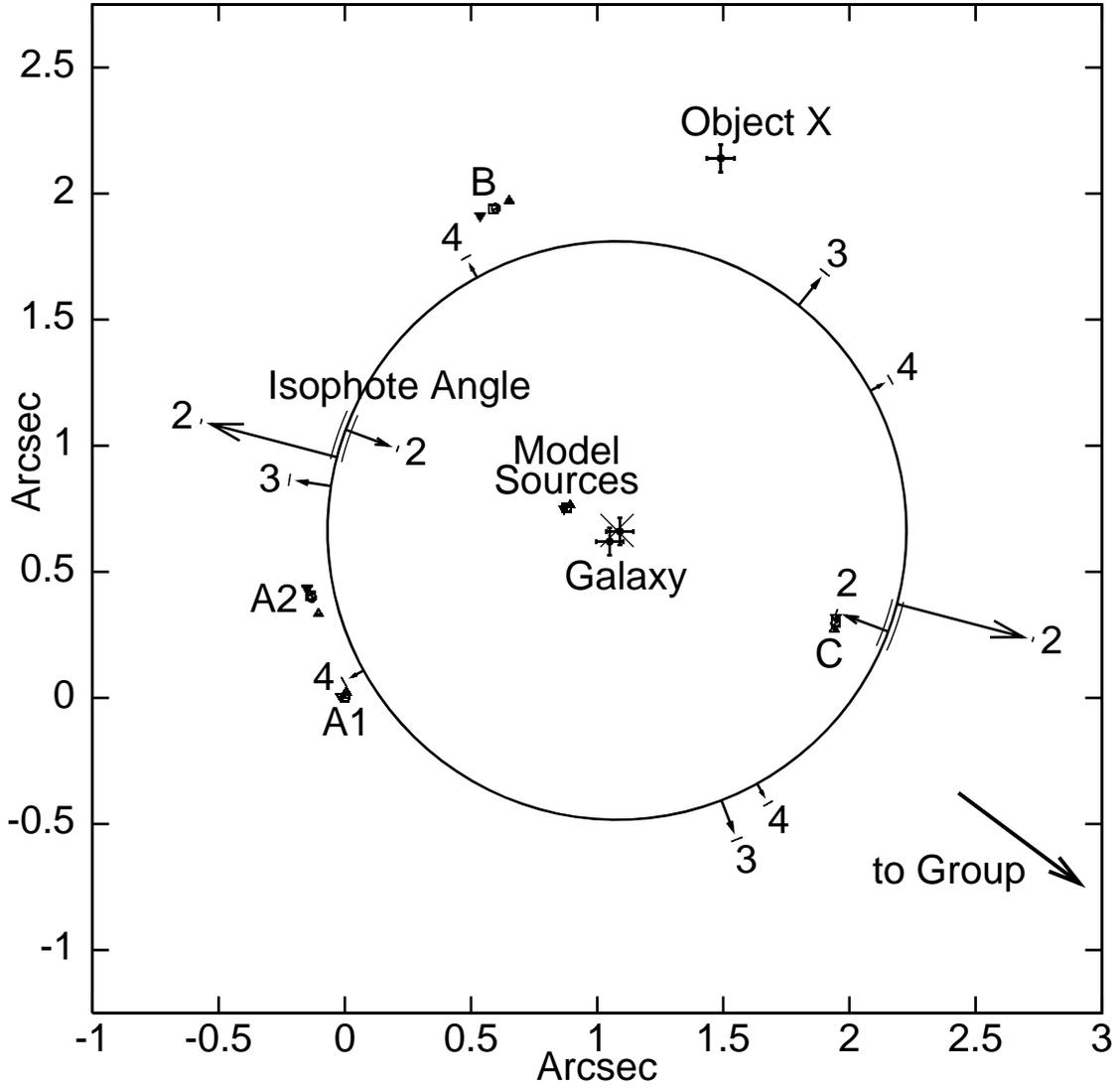}{6in}{0}{200}{200}{-350}{-100}
\caption{
Model $b + \AmvecWith{2} + \BmvecWith{2} + \AmvecWith{3} + \AmvecWith{4}$, fitted to
MG~J0414+0534.
The model parameters are illustrated graphically,
using the same conventions as in Figure~\protect\ref{fig:A3}.
}
\label{fig:A3_A4}
\end{figure}

The best-fit model of the lens potential with 11 parameters included
an $m=3$ external multipole and an $m=4$ external multipole. What are
the implications of this success for the mass distribution of the deflector?
To recapitulate,
the explicit form of the model potential in this case is:
\begin{eqnarray}
\Phi'(x,y) & = & b^2 
\left( \vstrut \right.
\rho - \left( (1+\rho)^2\AmvecWith{2}' + \frac{1}{(1+\rho)^2}\BmvecWith{2}' \right) \cdot
(\xhat\cos 2\theta + \yhat\sin 2\theta) - \\
& & (1+\rho)^3\AmvecWith{3}' \cdot (\xhat\cos 3\theta + \yhat\sin 3\theta) -
(1+\rho)^4\AmvecWith{4}' \cdot (\xhat\cos 4\theta + \yhat\sin 4\theta)
\left. \vstrut \right)
\nonumber
\end{eqnarray}
where $\theta$ is measured 
north of west about 
the center of mass $(g_{x},g_{y})$ 
as given by $\tan \theta = (y-g_y)/(x-g_x)$,
and the
radial parameter $\rho$ is given by
\begin{equation}
\rho = \left(  \sqrt{ (x-g_x)^2 + (y-g_y)^2} - b \right) / b.
\end{equation}
Table~\ref{tbl:best-fit-parameters} contains a list of the
best-fit parameters. In addition, Table~\ref{tbl:magnifications}
lists the image magnifications predicted
by this best-fit model (which may
be compared to the flux ratios in Table~\ref{tbl:positions}).
Figure~\ref{fig:residuals} shows both the
observed and modeled image locations for each of the sixteen
components of the VLBI map, along with the stat-width error
ellipses.

In this model, the center of mass of the mass
distribution interior to $r=b$ is located at
$\Delta\alpha = -1\arcsec.0788 \pm 0\arcsec.0020$,
$\Delta\delta = 0\arcsec.6635 \pm 0\arcsec.0012$
relative to the correlation center at component A1p.
Since the position of component Cp relative to the correlation center was
$\Delta\alpha = -1\arcsec.9454$, $\Delta\delta = 0\arcsec.3004$ 
(with position errors negligible compared to those of the model galaxy,
see Table~\ref{tbl:positions}) the model center of mass is 
$\Delta\alpha = 0\arcsec.8666 \pm 0\arcsec.0020$,
$\Delta\delta = 0\arcsec.3631 \pm 0\arcsec.0012$
relative to Cp.
This is in agreement with the optical
centroid of the lens galaxy as observed by
Falco, Leh\'{a}r \& Shapiro (1997)
even though the optical position was not used as a modeling constraint.
The observed lens galaxy position in $R$-band, relative to component C, was
$\Delta\alpha = 0\arcsec.90 \pm 0\arcsec.05$,
$\Delta\delta = 0\arcsec.32 \pm 0\arcsec.05$;
the observed position in $I$-band was
$\Delta\alpha = 0\arcsec.86 \pm 0\arcsec.05$,
$\Delta\delta = 0\arcsec.36 \pm 0\arcsec.05$.

As is apparent in Table~\ref{tbl:best-fit-parameters}
and Figure~\ref{fig:A3_A4},
the directions of mass excess implied by both
the internal and external quadrupoles ($\AmvecWith{2}'$ and
$\BmvecWith{2}'$) agree
with the direction of the optical isophotes ($71\arcdeg\pm 5\arcdeg$)
observed by Falco, Le\'{a}r \& Shapiro (1997).
Interestingly, one of the directions of mass deficit implied by the
external $m=4$ multipole, $\AmvecWith{4}'$, is also in agreement with the
optical isophote angle.
The eastern direction of the external octupole, $\AmvecWith{3}'$, is aligned with
the external quadrupole moment within $6\arcdeg$, 
and is within $10\arcdeg$ of the observed isophote angle, 
although in neither case do the formal confidence
regions overlap.
It is possible that the alignments
between the multipole angles and the optical isophotes are coincidences.
If the isophote angle were selected at random, the chance of agreement
with the direction of mass-excess indicated by $\AmvecWith{2}'$ 
would be $6\%$, given the quoted confidence ranges. 
The chance of agreement with the direction of either the mass excess
or mass deficit implied by $\AmvecWith{4}'$ would be 27\%.
The chance that $\AmvecWith{3}'$ and the isophote angles would
be as closely aligned as they are is 34\%.

Bearing this in mind, we entertain three speculations
regarding the origin of the
external multipoles in the best-fit model
$b+\AmvecWith{2}+\BmvecWith{2}+\AmvecWith{3}+\AmvecWith{4}$:
 
\begin{enumerate}

\item All of the external multipoles are attributable to the mass
distribution of the lens galaxy. This explains the alignments of
the various multipole
angles with the optical isophotes, but it implies that no external
perturber (i.e. neither
object X nor the group of galaxies to the southwest) contributes significantly
to the features of the potential we have modeled.
A singular isothermal elliptical potential would have 
a ratio of external to internal quadrupole of
$|\AmvecWith{2}|/|\BmvecWith{2}| = 3$,
which is consistent with the value $2.90\pm 0.17$ obtained for this model.
The ellipticity ($1-b/a$) of the isopotential contours near the ring radius for
the fitted model
is $\epsilon_\Phi = 2|\AmvecWith{2}'+\BmvecWith{2}'| = 0.120 \pm 0.002$.
In contrast the ellipticity of a singular isothermal elliptical
potential (SIEP) having an
isodensity ellipticity of $\epsilon_G = 0.20 \pm 0.02$
(equal to the mean ellipticity of the fitted isophotes 
of \cite{hst})
would be $\epsilon_\Phi = 0.07$.
However, the non-zero $\AmvecWith{3}'$ implies that the outer galactic halo is
asymmetric, with more mass concentrated near one end of the isophote axis than the other.
Furthermore, the magnitude of $\AmvecWith{4}'$ ($4.1\times 10^{-3}$)
is an order of magnitude larger than the value
it would assume for an singular 
isothermal elliptical potential ($6\times 10^{-4}$).
The direction of
$\AmvecWith{4}'$ implies that the mass distribution is box-like, rather than disk-like as for a SIEP.

\item $\AmvecWith{2}'$ and $\AmvecWith{3}'$
indicate an external perturbing mass to the east.
In this case
the alignments of all multipoles with the isophotes are accidental.
According to equation~\ref{eq:external-multipole-moment-of-PM-perturber},
a point mass with
an Einstein radius of $0.95\pm0.05$ arcseconds, located $3.2\pm 0.2$ arcseconds
away, would supply approximately
the appropriate values of $\AmvecWith{2}'$ and $\AmvecWith{3}'$, but would not
account for $\AmvecWith{4}'$.
No perturber of any kind is seen to the east in the optical images of \cite{hst}.

\item $\AmvecWith{3}'$ and $\AmvecWith{4}'$
indicate an external perturbing mass 
at about $-65\arcdeg \pm 4\arcdeg$
N of W. In this case, as above,
the alignments of all multipoles with the isophotes are accidental. 
A point mass $2.24\pm 0.24$ arcseconds away, with Einstein radius $0.56\pm 0.10$
arcseconds, would supply the proper $|\AmvecWith{3}|$ and $|\AmvecWith{4}|$, but
would also make a significant contribution to $\AmvecWith{2}$.
The residual component of $\AmvecWith{2}'$, which would presumably be due to the
lensing galaxy, has magnitude 0.06 and direction $1.5\arcdeg$ or $181.5\arcdeg$
north of west. This is somewhat problematic because the residual quadrupole does not
agree in angle with the optical isophotes, and its magnitude is uncomfortably large
($|\AmvecWith{2}'|/|\BmvecWith{2}'| = 3.8$).
However, there is an object seen in the optical image of \cite{hst} about
$5\arcsec$ away in the direction $-54\arcdeg$ N of W.

\end{enumerate}

None of these speculations is entirely satisfactory. The first speculation
attributes a
peculiar shape to the galactic halo; the second invokes an external perturber
that does not seem to be present; the third compels the galaxy to produce an
unusual quadrupole moment. We favor the first interpretation, because it
is hard to arrange for an external perturbation to produce $\AmvecWith{3}'$ without
ruining the suggestive alignment of $\AmvecWith{2}'$ with the isophotes, but admit
that this interpretation is debatable.
%
%

This ambiguity of interpretation illustrates both the appeal and
the frustration of the multipole-Taylor technique for modeling lens potentials.
The technique makes few preconceptions about the shape of the potential,
which in principle may lead to unanticipated discoveries about the mass
distribution of the deflector, including the shape of the halo (of which
little is presently known). However, precisely because of this mathematical
generality, there is no determinative way to correlate features of the potential
with observed astrophysical objects (the lens galaxy or external perturbers).

Finally, we comment on the unconstrainability of the radial distribution
of the mass monopole.
It is not terribly surprising that we were unable to usefully constrain
the parameter $f_{3}'$, the lowest-order parameter containing
information about the radial distribution.
A large change in $f_{3}'$ is needed
to cause a small change in the radial positions of images
near the ring radius.  By contrast, even small values of $\Amvec'$
or $\Bmvec'$ (for $m\geq 2$) cause shifts in the
radial positions of images near the ring radius.
In particular, varying $\Amvec'$ and $\Bmvec'$ while
leaving their sum unchanged affects the
radial image displacements, but has little effect on the
tangential image displacements (see
Table~\ref{tbl:first-derivatives}).
The radial image shifts caused by these multipole terms
depend on angular position, whereas those caused by $f_{3}'$
do not. However, for lenses such as MG~J0414+0534, in which there
are only images at a few angular locations,
the effects of $f_{3}'$ and of $\Amvec'$
or $\Bmvec'$ may compete, with large changes in $f_{3}'$
compensating for small changes in $\Amvec'$
or $\Bmvec'$.
For systems with arcs or rings of lensed emission,
information is available from a broader range of angles.
The optical arc visible in MG~J0414+0534 (\cite{hst})
may further constrain MG~J0414+0534's angular multipole moments,
especially if its location can be measured
with the same precision as the VLBA measurements
used in this paper.

That the radial profile parameter, $f_{3}'$, is so difficult to determine
is unfortunate. It would provide information on
how the angularly-averaged surface mass density
decreases with radius near the Einstein ring radius, which could
be used help choose a model value for $\sigma_0(b)$.
The quantity $\sigma_0(b)$ is not directly constrainable
from lensing, but it does affects the predicted time delays
between images, a topic discussed in the next section.

\subsection{Implications for the time delays}
\label{sec:implications-for-time-delays}

Each of the multiple images formed by a gravitational lens represents the source
object at a different moment in its history. This is because the propagation time from
source to observer is different for each image, due to the different path lengths and
Shapiro delays experienced by the ray bundles composing each image. In particular,
the time delay by which an image at $\rvec_{\alpha}$ lags that at $\rvec_{\beta}$ is,
as computed by Narayan \& Bartelmann (1996),
\begin{equation}
\label{eq:time-delay}
\Delta t_{\beta \alpha} = t(\rvec_{\alpha}) - t(\rvec_{\beta}) =
(1+z_{l}) \frac{D_{L}D_{S}}{D_{LS}}
\left(
\frac{1}{2} |\Del_{\rvec_{\alpha}}\Phi(\rvec_{\alpha})|^2 
                    - \frac{1}{2} |\Del_{\rvec_{\beta}}\Phi(\rvec_{\beta})|^2
             - \Phi(\rvec_{\alpha}) + \Phi(\rvec_{\beta}) 
\right)
\end{equation}
It is convenient to define a ``dimensionless time
delay'' $\Delta \tau_{\beta\alpha}$
which depends only on the modeled lens potential
and requires no values for redshifts or cosmological parameters,
\begin{equation}
\Delta \tau_{\beta \alpha}  = 
        \frac{1}{2} |\Del_{\rvec_{\alpha}}\Phi(\rvec_{\alpha})|^2 
      - \frac{1}{2} |\Del_{\rvec_{\beta}}\Phi(\rvec_{\beta})|^2
       - \Phi(\rvec_{\alpha}) + \Phi(\rvec_{\beta}) 
\end{equation}
and which is related to the time delay by a conversion factor depending
on the redshifts and cosmology,
\begin{equation}
\Delta t_{\beta\alpha}  = 
(1+z_l) \frac{D_L D_S}{D_{LS}} \Delta\tau_{\beta\alpha}.
\label{eq:delta-t-to-delta-tau}
\end{equation}

A measurement of the time delay between the flux variations
in corresponding images and
the lens redshift $z_{l}$, 
when combined with a model that predicts the dimensionless time delay,
thereby amounts to a measurement of the combination
of angular diameter distances $D_{L}D_{S}/D_{LS}$.
Since the relation between angular diameter distance
and redshift depends on the values of $H_{0}$,
$\Omega_{m}$, and $\Omega_{\Lambda}$,
%
%
these cosmological parameters can be thereby constrained.
The appeal of this well-known cosmological
probe is that it does not rely on the usual intermediate distance indicators
(\cite{refsdal}).

One problem with this idea arises from the mass-sheet degeneracy,
which was discussed in section~\ref{sec:degeneracies}.  When the
potential is transformed by the mass sheet degeneracy, the time delays
between components are likewise transformed.  Using the model
potential of equation~\ref{eq:final-form-of-potential}, 
the model dimensionless time delay,
$\Delta \tau'$, is related to the true dimensionless time delay,
$\Delta \tau$, by
\begin{equation}
\Delta \tau'_{\alpha\beta} = 
\frac{1}{2\left(1-\frac{\sigma_0(b)}{\sigcrit}\right)}
\Delta \tau_{\alpha\beta}
\end{equation}
where $\sigma_0(b)$ is the angularly-averaged surface mass density at
the Einstein ring radius.  Unless the mass-sheet degeneracy can be
resolved by determining $\sigma_0(b)/\sigcrit$ in some independent fashion
the time delays cannot be predicted,
although the ratios of time delays
between different image pairs can still be predicted.

Although MG~J0414+0534 has been extensively monitored at
radio wavelengths, no flux variations
have been observed that are large enough to permit an
accurate time delay measurement (\cite{moore}).
Nevertheless, this does not preclude radio or
optical detections of time delays in the future, so it
is important to understand how our models of MG~J0414+0534 constrain the time delays.
As discussed in the previous section, the best-fit model was
$b+\AmvecWith{2}+\BmvecWith{2}+\AmvecWith{3}+\AmvecWith{4}$. We used this model
to make our single best prediction for the time delays of MG~J0414+0534,
by computing the dimensionless time delays
between the 4 images of the brightest component (p).

The uncertainty in 
this particular model's predicted
time delay due to the uncertainty in the measured
image positions was estimated in the following manner.
We re-computed the time delay with
each parameter (one at a time) adjusted to its maximum and minimum
values allowed by the stat-width confidence limits, 
while minimizing the $\chi^2$ over all the other parameters.
The range between the highest and lowest time delays that were achieved
during these parameter-by-parameter adjustments is our estimate
for the uncertainty in the time delay for that particular model.

However, we must also take into account the larger uncertainty in
the predicted time delay caused by the uncertain choice of model.
To estimate this uncertainty, we
computed dimensionless time delays for a large subset of the models
that were discussed in the previous section.
For the 11-parameter and
12-parameter models, we included models which adequately fit the
observations when using the stat-width error estimates
($N_\sigma < 3$).
The $+\AmvecWith{3}+\MsummvecWith{4}+f_3$
and $+\AmvecWith{3}+\BmvecWith{4}+f_3$ models were excluded
because, for these models, the entire confidence range for $f'_3$ lies 
above $f'_3 > 2$, well within the
unphysical region $f'_3 > 1$.
Likewise, the $+\AmvecWith{3}+\Gvec_{22}$ model was not included
because of its unphysically large value of $|\Gvec'_{22}|$,
as discussed in section~\ref{sec:model-results}.
The results are shown in Table~\ref{tbl:dimensionless-time-delays}.
Since these models make somewhat different predictions, any attempt to
make a single prediction for the time delays must incorporate the
uncertainty associated with the selection of a single model.  Thus
we have enlarged the error spread in our best predictions
for the time delays to include
the whole range of time delays predicted by all these models.
The resulting predictions are:

\begin{eqnarray}
\Delta\tau_{B A}' 
= \frac{1}{2\left(1-\frac{\sigma_0(b)}{\sigcrit}\right)}
\Delta\tau_{B A}
& = & 1.828\times10^{-12} 
\, 
\leftParen 1 
        {}^{+}_{-}
        \underbrace{{}^{0.020}_{0.018}}_\formalerrors
        \pm \underbrace{0.014}_\AoneAtwodifference
        {}^{+}_{-}
        \underbrace{{}^{0.616}_{0.038}}_\whichmodel
 \rightParen,
\\
\Delta\tau_{A C}' 
= \frac{1}{2\left(1-\frac{\sigma_0(b)}{\sigcrit}\right)}
\Delta\tau_{A C}
& = & 1.042\times10^{-11} 
\, 
\leftParen 1 
        \pm \underbrace{0.017}_\formalerrors
        \pm \underbrace{0.002}_\AoneAtwodifference
        {}^{+}_{-}
        \underbrace{{}^{0.085}_{0.198}}_\whichmodel
 \rightParen,
\\
\Delta\tau_{B C}' 
= \frac{1}{2\left(1-\frac{\sigma_0(b)}{\sigcrit}\right)}
\Delta\tau_{B C}
& = & 1.225\times10^{-11} 
\, 
\leftParen 1 
        \pm \underbrace{0.017}_\formalerrors
        {}^{+}_{-}
        \underbrace{{}^{0.138}_{0.148}}_\whichmodel
 \rightParen.
\end{eqnarray}

Here the ``formal errors'' represent the uncertainty in the parameters
of the best-fit model, the ``A1-A2 difference'' is half the time delay
between images A1 and A2 (since the joint A1-A2 light curve would probably
be used to measure time delays), and ``which model'' refers to the uncertainty
due to choice of model. The factor 
$2 \left( 1-{\sigma_0(b)}/{\sigcrit} \right)$
represents the uncertainty
in the time delay due to the mass-sheet degeneracy, which can only be relieved
by obtaining a reliable value of 
${\sigma_0(b)}/{\sigcrit}$
from other observational or theoretical
sources. For a potential with the radial profile of an isothermal sphere, 
$2 \left( 1-{\sigma_0(b)}/{\sigcrit} \right) = 1$.

To express the time delay as an actual number of days, the conversion
factor of equation~\ref{eq:delta-t-to-delta-tau} must be used.  
For MG~J0414+0534, which has $z_l = 0.9584$ and $z_s = 2.639$,
this conversion factor takes the value $\Delta t / \Delta \tau =
6.794 \times 10^{12} \, h_{75}^{-1}$ days, assuming the universe has $\Omega_{m}=1$,
$\Omega_{\Lambda}=0$, and $H_{0} = 75h_{75}$ km/s/Mpc.
The values of this conversion factor
for some other choices of cosmological parameters are tabulated in
Table~\ref{tbl:conversion-factors}. 
In all cases, the ``filled-beam'' approximation was used to compute
the conversion factor, in which the universe is assumed to have a perfectly
smooth distribution of matter. The presence of clumpiness would require the
angular-diameter distances to be re-computed (see e.g. \cite{fukugita}).

A promising way to reduce the ``which model'' uncertainty is to measure the time delay
ratio $\Delta t_{AB} / \Delta t_{AC}$. Predictions for this ratio are
presented in Table~\ref{tbl:MMT-time-delay-ratios}
for various models. If this quantity is in accordance
with the prediction of our best-fit model
$b+\AmvecWith{2}+\BmvecWith{2}+\AmvecWith{3}+\AmvecWith{4}$,
and it can be measured to within
3\%, then it would exclude all the other models listed and the ``which model''
error would fall away. Even if the ratio could only be measured to within
18\%, it would exclude all but one other model, which would shrink the
``which model'' uncertainty in $\Delta t_{AB}$ to only 3.7\% (from 62\%).
In this scenario, 
and using the conversion factor (eq.~\ref{eq:delta-t-to-delta-tau})
for $\Omega_m = 1$, $\Omega_{\Lambda} = 0$,
we would predict 
\begin{equation}
\left.
\begin{array}{ccl}
H_0  & = & 75 \, {\rm km/s/Mpc} \,
\frac{12.42 \, \rm days}{\Delta t_{B A}} \, 
\leftParen 1 
        {}^{+}_{-}
        \underbrace{{}^{0.020}_{0.018}}_\formalerrors
        \pm \underbrace{0.014}_\AoneAtwodifference
        {}^{+}_{-}
        \underbrace{{}^{0.000}_{0.037}}_\whichmodel
 \rightParen
 \left[
        {2\left(1-\frac{\sigma_0(b)}{\sigcrit}\right)}
 \right]
\\
H_0  & = & 75  \, {\rm km/s/Mpc} \,
\frac{70.79 \, \rm days}{\Delta t_{A C}} \, 
\leftParen 1 
        \pm \underbrace{0.017}_\formalerrors
        \pm \underbrace{0.002}_\AoneAtwodifference
        {}^{+}_{-}
        \underbrace{{}^{0.000}_{0.111}}_\whichmodel
 \rightParen
 \left[
        {2\left(1-\frac{\sigma_0(b)}{\sigcrit}\right)}
 \right]
\\
H_0  & = & 75  \, {\rm km/s/Mpc} \,
\frac{83.23 \, \rm days}{\Delta t_{B C}} \, 
\leftParen 1 
        \pm \underbrace{0.017}_\formalerrors
        {}^{+}_{-}
        \underbrace{{}^{0.000}_{0.100}}_\whichmodel
 \rightParen
 \left[
        {2\left(1-\frac{\sigma_0(b)}{\sigcrit}\right)}
 \right]
\end{array}
\right\}
\rule{0.5em}{0ex}
\parbox{0.18\textwidth}{
\setlength{\baselineskip}{0.9\baselineskip}
Assuming the time delay ratio is
measured to within $\sim$18\% and
agrees with our best-fit model.}
\end{equation}

\section{Conclusions}

Upon first seeing the rich sub-structure in each VLBI image of MG~J0414+0534, we were
hopeful that such a large body of precise constraints on the modeling potential would
lead to tight constraints on the mass distribution of the deflector and the
predicted time delays between images. We hoped that the mathematical generality
of the multipole-Taylor expansion (with slight modifications) would allow us to draw
such conclusions without contamination from (perhaps faulty) astrophysical preconceptions.
These hopes were only partly fulfilled.

Once the best-fit parameters in the expansion are determined, it is
difficult to know from which astrophysical source they arise. For example,
our best-fit model 
$b+\AmvecWith{2}+\BmvecWith{2}+\AmvecWith{3}+\AmvecWith{4}$
seems to imply that either the mass distribution in the lensing galaxy is asymmetric
and somewhat quadrangular (boxy), or else that an unseen external perturber is partly
responsible for the light deflection (as discussed in 
section~\ref{sec:discussion-mass}). It is possible, however, that
the values of the best-fit model parameters
may guide future interpretations of observations for this system, by
indicating the possible directions of perturbing masses.

Predictions of time delays, while quite well-constrained for any particular
model potential, are limited by the uncertainty in selecting one of several viable
models. In other words, the predictions are not limited by the precision
of the positional constraints, but rather by the ability to satisfy those
constraints with several alternative truncations of the multipole-Taylor series.
In the case
of MG~J0414+0534, we found that one way to relieve this crucial source
of systematic error is to measure ratios of time delays, which are predicted
to have different values by different models.

Despite these limitations, which afflict all lens modeling techniques
to date,
the multipole-Taylor expansion does indeed seem to be an appropriate form for
multiple-image gravitational lenses such as MG~J0414+0534. 
The simplest
three-term truncation reproduces the observed image configuration far better
than previously-used simpler models, and successively higher terms improve
the fit by incrementally smaller amounts.
The radial profile parameter $f'_3$ could not be constrained,
but this is likely to be the case for any point-modeling scheme applied
to quadruple-image lenses.

We believe the multipole-Taylor expansion could be usefully applied to
other lenses in which the constraints occur at locations
close to the Einstein radius, and the angular variation of the potential is
expected to be fairly smooth. One serious problem with the efforts to date
to determine $H_0$ and other cosmological parameters by measuring
time delays is that each known gravitational lens has usually been modeled
in an individual and idiosyncratic manner. The multipole-Taylor expansion
is one candidate for a very general modeling technique that could
be applied to all the time-delay lenses, so that the results of these
efforts could be sensibly combined.

\acknowledgments

We thank Paul Schechter for many valuable discussions concerning 
lens modeling in general and this analysis of MG~J0414+0534
in particular. J. N. Winn thanks the Fannie and John Hertz
Foundation for financial support. This work was supported by
grant AST96-17028 from the National Science Foundation.

\clearpage

\begin{table}
\noindent
\begin{center} 
\begin{tabular}{crrccccr} \tableline
&
\multicolumn{2}{c}{Component location} &
\multicolumn{3}{c}{Centroid errors} &
Peak flux &
Integral flux  
\\
& $x = -\mbox{R.A.}$ & $y = \mbox{Dec.}$ & $\sigma_x$ & $\sigma_y$ & $\sigma_{xy}^2$ & (mJy/beam)
& \multicolumn{1}{c}{(mJy)}
\\
& (mas) & (mas) & (mas) & (mas) & (mas$^{2}$) &  & 
\\ \tableline \tableline	 
  A1p  &     0.0144 &    0.3818 &   0.0018  & 0.0037 & -2.33e-6  & $108.1\pm0.2$ & $149.1\pm 0.4$  \\  
  A2p  &  -134.0714 &  405.9972 &   0.0025  & 0.0062 & -7.09e-6  & $88.0\pm0.3$ & $125.9\pm0.5$ \\      
   Bp  &   588.6037 & 1938.3514 &   0.0056  & 0.0089 &  8.09e-6  & $  33.2 \pm 0.2 $ & $  43.4 \pm 0.4 $ \\       
   Cp  &  1945.3597 &  300.4118 &   0.0061  & 0.0156 &  6.06e-6  & $  20.5 \pm 0.2 $ & $  22.5 \pm 0.3 $ \\ \tableline
  A1q  &     1.9438 &    1.6872 &   0.0044  & 0.0103 & -9.76e-6  & $  39.2 \pm 0.2 $ & $  48.8 \pm 0.4 $  \\  
  A2q  &  -130.5108 &  397.7602 &   0.0075  & 0.0195 & -7.12e-5  & $  28.1 \pm 0.3 $ & $  39.2 \pm 0.5 $  \\ 
   Bq  &   598.4155 & 1943.2536 &   0.0215  & 0.0263 &  1.53e-4  & $  10.7 \pm 0.2 $ & $  17.5 \pm 0.4 $  \\ 
   Cq  &  1944.7253 &  296.2398 &   0.0203  & 0.0401 &  2.22e-4  & $   6.5 \pm 0.2 $ & $   6.6 \pm 0.3 $  \\ \tableline 	
  A1r  &     6.7992 &   21.2226 &   0.0298  & 0.0563 & -1.30e-3   & $   9.2 \pm 0.2 $ & $  20.2 \pm 0.6 $  \\  
  A2r  &  -104.5880 &  332.3330 &   0.0243  & 0.0770 & -1.31e-3   & $  10.1 \pm 0.2 $ & $  22.7 \pm 0.7 $  \\      
   Br  &   652.1233 & 1970.6440 &   0.0356  & 0.0626 &  1.69e-4    & $   5.2 \pm 0.2 $ & $   7.1 \pm 0.4 $  \\  
   Cr  &  1939.5305 &  271.3034 &   0.0365  & 0.1004 &  2.01e-6   & $   2.7 \pm 0.2 $ & $   2.5 \pm 0.2 $  \\ \tableline 
  A1s  &   -16.1756 &    6.1174 &   0.0854  & 0.1362 & -1.13e-2   & $   6.6 \pm 0.1 $ & $  61.6 \pm 1.2 $  \\ 
  A2s  &  -149.3272 &  432.3058 &   0.0236  & 0.1334 & -9.98e-4   & $   8.4 \pm 0.2 $ & $  49.7 \pm 1.2 $  \\  
   Bs  &   536.6837 & 1912.7346 &   0.0978  & 0.0561 &  3.02e-3   & $   4.2 \pm 0.1 $ & $  17.5 \pm 0.7 $  \\   
   Cs  &  1945.9419 &  319.7664 &   0.0617  & 0.1036 & -4.45e-3  & $   4.4 \pm 0.2 $ & $  15.2 \pm 0.6 $  \\
\tableline
\end{tabular}
\end{center}
\caption{
Centroid locations and flux densities of the elliptical Gaussian fits to
the components of MG~J0414+0534, derived from the VLBA observations
described by Trotter (1998).
The positions are relative to the
correlation and fringe-fitting center at A1.
The $x$-coordinate increases to the west
and the $y$-coordinate increases to the north.
The errors on the centroid positions are standard deviations on
the $x$- and $y$-positions due to the thermal noise in the map, and
the corresponding correlation between the $x$- and $y$-position
errors.
The flux errors are standard deviations due to the thermal noise in
the map.  These error bars do not include the effect of deconvolution
error.
}
\label{tbl:positions}
\end{table}

\begin{table}
\noindent
\begin{center}
\begin{tabular}{p{0.11\textwidth}p{0.11\textwidth}p{0.10\textwidth}p{0.56\textwidth}} \tableline
{Multipole moment $m$}   & {Radial dependence} & {Parameter} & {Significance} \\ \tableline \tableline
\nodata		& \nodata	& $g_x$, $g_y$		& Location of origin. \\
$m=0$		& $t=0$		& $f_0$			& Constant offset (un\-con\-strain\-a\-ble). \\
$m=0$		& $t=1$		& $b$			& Einstein ring radius. Sensitive to total mass within ring. \\
$m=0$		& $t=2$		& $f_2$			& Sets surface mass density at ring, $\sigma_0(b)$ 
                                                          (un\-con\-strain\-a\-ble due to mass-sheet degeneracy). \\
$m=0$		& $t\geq 3$	& $f_t$			& Depends on $\sigma_0(r)$ and its derivatives near $r=b$. 
                                                          For $t=3$, gives falloff of surface mass density with radius, 
                                                          evaluated at the ring radius. 
                                                          See Eq.~44. \\
$m=1$		& exterior	& $\AmvecWith{1}$	        & Dipole moment of mass exterior to ring 
                                                          (un\-con\-strain\-a\-ble due to prismatic degeneracy). \\ 
$m\geq 2$	& exterior	& $\Amvec$		& Sensitive to multipole moment of mass exterior to ring. \\
$m=1$		& interior	& $\BmvecWith{1}$		& Sets COM of mass interior to ring. (The COM may
                                                          be placed at the origin by setting $\BmvecWith{1}=0$.) \\
$m\geq 2$	& interior	& $\Bmvec$		& Sensitive to multipole moment of mass interior to ring. \\
$m\geq 1$	& $t\geq 2$	& $\Gmtvec$		& Depends on $\sigvec_{m}(r)$ and its derivatives near $r=b$.\\
$m\geq 2$	& $t=0$		&$\Msummvec$		& $\Msummvec = \Amvec + \Bmvec$, sum of external and 
                                                          internal mass multipoles.  (This parameter is from the 
                                                          pa\-ram\-e\-ter\-i\-za\-tion    
                                                          in equation~7, rather 
                                                          than equation~47). \\ \tableline 
\end{tabular}
\end{center}
\caption{Physical significance of the parameters and terms in
the modified multipole-Taylor series, equation~47.
The first column lists the multipole moment $m$.  The second column describes the
radial dependence (the exponent $t$ of $\rho^t$, or whether the
term is sensitive to mass interior or exterior to $r=b$).
The third column lists the name of the parameter. 
The last column describes the physical significance of the term.
The parameters $f_0$, $f_2$, $\AmvecWith{1}$, and $\BmvecWith{1}$ cannot
be constrained from lensing, as explained in section~3.4.
We set these parameters equal to zero during the model-fitting procedure.
\label{tbl:physical-significance}
}
\end{table}

\begin{table}
{
\begin{center}
\begin{tabular}{cccp{0.24\textwidth}} \tableline
$m$	& $t$		
			& $\frac{1}{b} \Del_{\rvec}\Phi'$	
			& Image Displacements \\ \tableline \tableline
$m=0$	& $t=0$		
			& $0$				
			& None \\ \tableline
$m=0$	& $t=1$		
			& $\rhat$			
			& Radial \\ \tableline
$m=0$	& $t\geq2$ 	
			& $\rhat \left\{ \frac{1}{(t-1)!} \rho^{(t-1)} \right\}	\left\{ \ft' \right\}$			
			& Small, radial \\ \tableline
$m\geq1$& exterior	
			& $\begin{array}[t]{c}
				 \rhat \left\{ - m (1+\rho)^{(m-1)} \right\} 
					\left\{ \Aone \right\} \\
				+ \thetahat \left\{ m (1+\rho)^{(m-1)} \right\} 
					\left\{ \Atwo \right\}
			   \end{array}$
			& Radial and tangential \\ \tableline	
$m\geq1$& interior	
			& $\begin{array}[t]{c}
				 \rhat \left\{ \frac{m}{(1+\rho)^{(m+1)}} \right\} 
					\left\{ \Bone \right\} \\
				+ \thetahat \left\{ \frac{m}{(1+\rho)^{(m+1)}} \right\} 
					\left\{ \Btwo \right\}
			   \end{array}$
			& Radial and tangential \\ \tableline	
$m\geq1$& $t\geq2$	
			& $\begin{array}[t]{c}
				 \rhat \left\{  \frac{\rho^{(t-1)}}{(t-1)!} \right\} 
					\left\{ \Gone \right\} \\
				+ \thetahat \left\{ -\frac{m}{t!} \frac{\rho^{t}}{(1+\rho)} \right\} 
					\left\{ \Gtwo \right\}
			   \end{array}$
			& Small, predominantly radial \\ \tableline \tableline
$m\geq1$& $t=0$		
			& $\thetahat \left\{ \frac{m}{1+\rho} \right\} \left\{ \Msumtwo \right\} $
			& Tangential, due to o\-ver\-all strength of mul\-ti\-pole
                          $\Msummvec'=\Amvec'+\Bmvec'$ \\ \tableline
$m\geq1$& $t=1$		
			& $\begin{array}[t]{c}
				 \rhat \left\{ - m \right\} 
					\left\{ \Mdiffone \right\} \\
				 + \thetahat \left\{ \frac{\rho m^2}{1+\rho} \right\} 
					\left\{ \Mdifftwo \right\}
			   \end{array}$
			& Predominantly ra\-di\-al.  Af\-fect\-ed by bal\-ance be\-tween
                          ex\-ter\-nal and in\-ter\-nal mul\-ti\-pole  $\Mdiffmvec'=\Amvec'-\Bmvec'$
                          \\ \tableline 
\end{tabular}
\end{center}
}
\caption{
The first derivatives of the terms in our parameterization of
the lens potential. 
These are needed to calculate the displacement of the lens images from
the source positions (eq.~3).  The upper section
of the table lists the five functional forms needed to represent the
terms in the modified multipole-Taylor expansion of the potential
(eq.~47).  The lower section of the
table lists functional forms for terms in the original expansion
(eq.~7).  The identifications $m$ and
$t$ are as in Table~2.
The model parameters are written in Cartesian form, 
e.g. $\Amvec' =  \xhat \Amc'  + \yhat \Ams' $.
\label{tbl:first-derivatives}
}
\end{table}

\begin{table}
\begin{center}
\begin{tabular}{cccc} \tableline
$m$	& $t$		& $\lDel_{\rvec}\Del_{\rvec}\Phi'$ \\ \tableline \tableline
$m=0$	& $t=0$		& 0 \\ \tableline
$m=0$	& $t=1$		& $\thetahat\thetahat \left\{  \frac{1}{1+\rho}  \right\}$ \\ \tableline
$m=0$	& $t\geq2$ 	& $\begin{array}[t]{c}
				\rhat\rhat \left\{ \frac{1}{(t-2)!} \rho^{(t-2)} \right\}
					\left\{ \ft' \right\} \\
				+ \thetahat\thetahat \left\{ 
							\frac{1}{(t-1)!} \frac{\rho^{(t-1)}}{(1+\rho)}
						 \right\}
					\left\{ \ft' \right\}
			   \end{array}$ \\ \tableline
$m\geq1$& exterior	& $\begin{array}[t]{c}
				 \rhat\rhat 
					\left\{ -m(m-1)(1+\rho)^{(m-2)}	\right\} 
					\left\{ \Aone \right\} \\
				+ \thetahat\thetahat
					\left\{  m(m-1)(1+\rho)^{(m-2)} 	\right\} 
					\left\{ \Aone \right\} \\
				+ (\rhat\thetahat + \thetahat\rhat)
					\left\{  m(m-1)(1+\rho)^{(m-2)}	\right\} 
					\left\{ \Atwo \right\}
			   \end{array}$  \\ \tableline	
$m\geq1$& interior	& $\begin{array}[t]{c}
				 \rhat\rhat 
					\left\{ - \frac{m(m+1)}{(1+\rho)^{(m+2)}}	\right\} 
					\left\{ \Bone \right\} \\
				+ \thetahat\thetahat
					\left\{  \frac{m(m+1)}{(1+\rho)^{(m+2)}}	\right\} 
					\left\{ \Bone \right\} \\
				+ (\rhat\thetahat + \thetahat\rhat)
					\left\{  -\frac{m(m+1)}{(1+\rho)^{(m+2)}}	\right\} 
					\left\{ \Btwo \right\}
			   \end{array}$ \\ \tableline	
$m\geq1$& $t\geq2$	& $\begin{array}[t]{c}
				 \rhat\rhat 
					\left\{ \frac{\rho^{(t-2)}}{(t-2)!} \right\} 
					\left\{ \Gone \right\} \\
				+ \thetahat\thetahat
					\left\{ \frac{1}{t!} \frac{\rho^{(t-1)}}{(1+\rho)^2} 
						(t - (m^2-t)\rho)	\right\} 
					\left\{ \Gone \right\} \\
				+ (\rhat\thetahat + \thetahat\rhat)
					\left\{ - \frac{m}{t!} \frac{\rho^{(t-1)}}{(1+\rho)^2} 
						(t + (t-1)\rho)	\right\} 
					\left\{ \Gtwo \right\}
			   \end{array}$ \\ \tableline \tableline
$m\geq1$& $t=0$		& $\begin{array}[t]{c}
				 \thetahat\thetahat
					\left\{ \frac{m^2}{(1+\rho)^2} \right\} 
					\left\{ \Msumone \right\} \\
				+ (\rhat\thetahat + \thetahat\rhat)
					\left\{ - \frac{m}{(1+\rho)^2} \right\} 
					\left\{ \Msumtwo \right\}
			   \end{array}$ \\ \tableline
$m\geq1$& $t=1$		& $\begin{array}[t]{c}
				 \thetahat\thetahat
					\left\{ - \frac{m}{(1+\rho)^2} 
						\left(1 - (m^2-1)\rho\right)	\right\} 
					\left\{ \Mdiffone \right\} \\
				+ (\rhat\thetahat + \thetahat\rhat)
					\left\{ \frac{m^2}{(1+\rho)^2} \right\} 
					\left\{ \Mdifftwo \right\}
			   \end{array}$ \\ \tableline 
\end{tabular}
\end{center}
\caption{
The second derivatives of the terms in our parameterization of the
lens potential. These are needed for calculation of the magnification
matrix.  The notation is that of
Table~3.  The magnitude of the effect of
each term on the inverse magnification matrix
(eq.~55), may be seen by
considering the exponent of $\rho$.
\label{tbl:second-derivatives}
}
\end{table}

\begin{table}
\noindent
\begin{center}
\begin{tabular}{p{0.28\textwidth}cccc} \tableline 
Model   
        & $\Phi(\rvec)$
        & $\frac{\sigma_0(\bE)}{\sigcrit}$
        & $1-\kappa 
                = \frac{1}{2\left( 1-\frac{\sigma_0(\bE)}{\sigcrit} \right)} $
        & $\ftWith{3}' $
\\ \tableline \tableline
point mass 
        & $ \bE^2 \ln r $
        & 0
        & $1/2$
        & 1
\\ \tableline
singular isothermal sphere
        & $ \bE r $
        & $1/2$
        & 1
        & 0
\\ \tableline
mass sheet with $\sigma = \sigcrit$
        & $ \frac{1}{2} r^2 $
        & 1
        & $\infty$
        & $-1$
\\ \tableline
power law & & & & \\
\begin{tabular}{cl@{}} $\alpha \rightarrow 0$,  & point mass \\
                   $\alpha = 1/2$,              & isothermal \\
                   $\alpha \rightarrow 1$,      & mass sheet \end{tabular}
        & $ \frac{\bE^2}{2\alpha} \left( \frac{r}{\bE} \right)^{2\alpha} $
        & $\alpha$
        & $\frac{1}{2(1-\alpha)}$
        & $1 - 2\alpha$
\\ \tableline
power law with core radius
        & $ \frac{s^2 + \bE^2}{2\alpha} 
                \left( \frac{s^2+r^2}{s^2+\bE^2} \right)^\alpha $
        & $ \frac{s^2 + \alpha \bE^2}{s^2 + \bE^2} $
        & $ \frac{1}{2(1-\alpha)} \frac{s^2 + \bE^2}{\bE^2} $
        & $ \frac{ (1-2\alpha) \bE^2 - 3s^2 }{s^2 + \bE^2} $
\\ \tableline 
\end{tabular}
\end{center}
\caption{
Comparison of several physically motivated monopole profiles.
The first column is the lens potential.
The second column is the surface mass density at the ring
radius. If a lens with such a profile were fitted with the
multipole-Taylor model with $f_2'$ set to zero (which was
the choice for $f_2'$ we employed) then the
resulting scale factor $1-\kappa$ is listed in the third column.
The fourth column lists the value of $f'_3$, the lowest-order
monopole parameter that is sensitive to the radial mass profile.  
Note that if $f'_3 > 1$, the surface mass density
increases with radius near the ring radius.
\label{tbl:comparison-to-other-models}
}
\end{table}

\begin{table}
\noindent
\begin{center} {\footnotesize \renewcommand{\arraystretch}{0.75}
\begin{tabular}{lccllllll} \tableline
\tstrut & & & \multicolumn{6}{c}{Method for estimating positional errors} \\ \cline{4-9}
\tstrut & & & \multicolumn{2}{c}{fit-size} & \multicolumn{2}{c}{statistical} & \multicolumn{2}{c}{stat-width} \\
        & & & \multicolumn{2}{c}{(an upper limit)} & \multicolumn{2}{c}{(a lower limit)} & &  \\ 
Model & $N$ & $N_{DOF}$ & 
\multicolumn{1}{c}{$\chi^2$} & \multicolumn{1}{c}{$N_{\sigma}$} & 
\multicolumn{1}{c}{$\chi^2$} & \multicolumn{1}{c}{$N_{\sigma}$} & 
\multicolumn{1}{c}{$\chi^2$} & \multicolumn{1}{c}{$N_{\sigma}$} \\ 
\tableline \tableline  \tstrut
%
SIEP                          &  5   & 19  &   2.357e4 & 5.402e3   &  1.297e8 & 2.976e7   &  3.667e5 & 8.412e4  \\	          
SIS+XS ($b+\AmvecWith{2}$)      &  5   & 19  &   1.680e4 & 3.851e3   &  1.078e8 & 2.474e7   &  1.910e5 & 4.382e4  \\	          
PM+XS                         &  5   & 19  &   1.448e4 & 3.317e3   &  1.012e8 & 2.321e7   &  2.080e5 & 4.771e4  \\ \tableline      	  
\multicolumn{9}{c}{  } \\ \tableline      \tstrut
$b+\AmvecWith{2}+\BmvecWith{2}$   &  7   & 17  &   2.485e2 & 5.615e1   &  1.636e5 & 3.967e4   &  3.996e3 & 9.651e2  \\ \tableline      	  
\multicolumn{9}{c}{  } \\ \tableline      \tstrut
$+\BmvecWith{3}$                &  9   & 15  &   1.356e2 & 3.113e1   &  1.143e5 & 2.950e4   &  2.022e3 & 5.181e2  \\	          
$+\Gvec_{22}$                 &  9   & 15  &   1.757e2 & 4.149e1   &  9.875e4 & 2.549e4   &  1.357e3 & 3.464e2  \\	          
$+\ftWith{3}$                 &  8   & 16  &   1.108e2 & 2.371e1   &  1.040e5 & 2.601e4   &  1.438e3 & 3.554e2  \\	          
$+\Gvec_{12}$                 &  9   & 15  &   1.400e2 & 3.227e1   &  1.430e5 & 3.691e4   &  8.457e2 & 2.145e2  \\                
$+\AmvecWith{4}$                &  9   & 15  &   1.011e2 & 2.224e1   &  7.958e4 & 2.054e4   &  7.493e2 & 1.896e2  \\	          
$+\MsummvecTab{4}$            &  9   & 15  &   1.235e2 & 2.801e1   &  8.522e4 & 2.200e4   &  3.913e2 & 9.717e1  \\	          
$+\BmvecWith{4}$                &  9   & 15  &   4.168e1 & 6.889     &  5.551e4 & 1.433e4   &  1.983e2 & 4.733e1  \\	          
$+\AmvecWith{3}$                &  9   & 15  &   1.619e1 & 0.308     &  1.831e4 & 4.723e3   &  1.351e2 & 3.100e1  \\	          
$+\MsummvecTab{3}$            &  9   & 15  &   8.783   & -1.605    &  1.603e4 & 4.135e3   &  7.038e1 & 1.430e1  \\ \tableline         
\multicolumn{9}{c}{ } \\ \tableline      \tstrut
$+\AmvecWith{4}+\BmvecWith{4}$           & 11   & 13  &   1.917e1 &  1.712    &  3.370e4 & 9.344e3   &  1.297e2 &  3.235e1 \\ \tableline      \tstrut
$+\MsummvecTab{3}+\ftWith{3}$        & 10   & 14  &   8.113   & -1.573    &  1.596e4 & 4.261e3   & 6.284e1  &  1.305e1  \\               
$+\MsummvecTab{3}+\BmvecWith{4}$       & 11   & 13  &   5.087   & -2.195    &  6.435e3 & 1.781e3   & 5.647e1  &  1.206e1  \\               
$+\MsummvecTab{3}+\MsummvecTab{4}$   & 11   & 13  &   3.795   & -2.553    &  6.130e3 & 1.697e3   & 2.862e1  &  4.331    \\               
$+\MsummvecTab{3}+\AmvecWith{4}$       & 11   & 13  &   2.459   & -2.924    &  3.998e3 & 1.105e3   & 1.239e1  & -0.1692   \\               
$+\MsummvecTab{3}+\Gvec_{22}$        & 11   & 13  &   3.285   & -2.694    &  8.421e3 & 2.332e3   & 1.179e1  & -0.3348   \\               
$+\MsummvecTab{3}+\Gvec_{12}$        & 11   & 13  &   2.527   & -2.905    &  7.794e3 & 2.158e3   & 1.145e1  & -0.4291   \\ \tableline      \tstrut
$+\AmvecWith{3}+\MsummvecTab{4}$       & 11   & 13  &   9.435   & -0.989    &  1.223e4 & 3.389e3   & 5.506e1  &  1.167e1  \\               
$+\AmvecWith{3}+\BmvecWith{4}$           & 11   & 13  &   7.374   & -1.560    &  1.050e4 & 2.908e3   &  4.732e1 &  9.519   \\	          
$+\AmvecWith{3}+\ftWith{3}$            & 10   & 14  &   9.144   & -1.298    &  1.631e4 & 4.356e3   &  4.183e1 &  7.439   \\		  
$+\AmvecWith{3}+\Gvec_{12}$            & 11   & 13  &   7.254   & -1.594    &  1.528e4 & 4.235e3   &  3.612e1 &  6.413   \\                
$+\AmvecWith{3}+\BmvecWith{3}$           & 11   & 13  &   1.654   & -3.147    &  4.176e3 & 1.155e3   &  8.960   & -1.120   \\	          
$+\AmvecWith{3}+\Gvec_{22}$            & 11   & 13  &   4.783   & -2.279    &  5.986e3 & 1.657e3   &  8.387   & -1.279   \\	          
$+\AmvecWith{3}+\AmvecWith{4}$           & 11   & 13  &   1.257   & -3.257    &  3.863e3 & 1.068e3   &  6.073   & -1.921   \\ \tableline          
\multicolumn{9}{c}{ } \\ \tableline      \tstrut
$+\AmvecWith{4} + \BmvecWith{4} + \ftWith{3}$       & 12   & 12  &   1.814e1 & 1.773     &  2.871e4 & 8.285e3   &  8.107e1 & 1.994e1  \\ \tableline      \tstrut
$+\MsummvecTab{3}+\BmvecWith{4}+\ftWith{3}$       &12    & 12  &   4.889   & -2.053    &  6.342e3 & 1.827e3   &  3.179e1 &  5.711   \\                
$+\MsummvecTab{3}+\MsummvecTab{4}+\ftWith{3}$   &12    & 12  &   3.791   & -2.370    &  5.948e3 & 1.714e3   &  2.455e1 &  3.622   \\                
$+\MsummvecTab{3}+\AmvecWith{4}+\ftWith{3}$       &12    & 12  &   2.321   & -2.794    &  3.798e3 & 1.093e3   &  1.234e1 &  0.099   \\                
$+\MsummvecTab{3}+\Gvec_{22}+\ftWith{3}$        &12    & 12  &   1.364   & -3.070    &  2.795e3 & 8.033e2   &  9.373   & -0.758   \\                
$+\MsummvecTab{3}+\Gvec_{12}+\ftWith{3}$        &12    & 12  &   2.203   & -2.828    &  5.505e3 & 1.586e3   &  8.495   & -1.012   \\ \tableline      \tstrut
$+\AmvecWith{3}+\Gvec_{12}+\ftWith{3}$            &12    & 12  &   6.160   & -1.686    &  1.445e4 & 4.169e3   &  3.431e1 &  6.439   \\                
$+\AmvecWith{3}+\MsummvecTab{4}+\ftWith{3}$       &12    & 12  &   3.376   & -2.490    &  7.549e3 & 2.176e3   &  1.696e1 &  1.431   \\                
$+\AmvecWith{3}+\BmvecWith{3}+\ftWith{3}$           & 12   & 12  &   1.571   & -3.011    &  3.989e3 & 1.148e3   &  8.193   & -1.099   \\	          
$+\AmvecWith{3}+\Gvec_{22}+\ftWith{3}$            &12    & 12  &   3.163   & -2.551    &  4.895e3 & 1.410e3   &  7.334   & -1.347   \\                
$+\AmvecWith{3}+\BmvecWith{4}+\ftWith{3}$           & 12   & 12  &   2.000   & -2.887    &  4.914e3 & 1.415e3   &  6.157   & -1.687   \\		  
$+\AmvecWith{3}+\AmvecWith{4}+\ftWith{3}$           & 12   & 12  &   0.720   & -3.256    &  2.306e3 & 6.622e2   &  2.721   & -2.679   \\ \tableline          
\end{tabular}
}
\renewcommand{\arraystretch}{1.0}
\end{center}\vspace{-7pt}
\caption{
Results of fitting multipole-Taylor models to MG~J0414+0534.
For each model, the number of parameters $N$ and number of degrees
of freedom $N_{DOF}$ are listed, along with the minimum value
of $\chi^2$ obtained and the number of standard deviations
$N_\sigma$ separating the model and constraints. Models
with negative $N_\sigma$ oversatisfy the positional constraints.
Three different estimates for the positional errors were used
and are reported separately (as discussed in section~4.1).
The models are named as described in section~4.2.
\label{tbl:results}
}
\end{table}

\begin{table}
\begin{center}

\renewcommand{\arraystretch}{0.95}
\begin{tabular}{ll@{}l} \tableline
deflector positions: & & \\
\multicolumn{1}{p{2.0in}}{ (W and N of the correlation
center at A1)} 
& $\begin{array}[t]{@{}l@{}} g_x = 1.0788 \pm {0.0020} \\ 
                    g_y = 0.6635 \pm {0.0012} \end{array}$
& $\begin{array}[t]{@{}r@{}} \mbox{arcseconds} \\ 
                    \mbox{arcseconds} \end{array}$
\\ \tableline
ring radius:
& $\bE = 1.1474^{+0.0025}_{-0.0026} $
& arcseconds
\\ \tableline
internal quadrupole:    & $\BmAmpWith{2}'= 0.01542^{+0.00089}_{-0.00085}$ & \\
$\BmvecWith{2}' = \BmAmpWith{2}' ( \xhat \cos 2 \BmAngWith{2}
			         +\yhat \sin 2 \BmAngWith{2} ) $
& $\begin{array}[t]{@{}c@{}l@{}} 2\BmAngWith{2} & =-0.713^{+0.028}_{-0.028} \\ 
 		       \BmAngWith{2} & = 69.57^{+0.81}_{-0.80} \end{array}$

& $\begin{array}[t]{@{}l} \mbox{radians N of W} \\
		       \mbox{degrees E of N} \end{array}$
\\ \tableline
external quadrupole:    & $\AmAmpWith{2}'= 0.04478^{+0.00034}_{-0.00033}$ & \\
$\AmvecWith{2}' = \AmAmpWith{2}' ( \xhat \cos 2 \AmAngWith{2}
			         +\yhat \sin 2 \AmAngWith{2} ) $
& $\begin{array}[t]{@{}c@{}l@{}} 2\AmAngWith{2} & =-0.513^{+0.013}_{-0.012} \\ 
 		       \AmAngWith{2} & = 75.29^{+0.36}_{-0.35} \end{array}$

& $\begin{array}[t]{@{}l} \mbox{radians N of W} \\
		       \mbox{degrees E of N} \end{array}$
\\ \tableline
external $m=3$ multipole: &$\AmAmpWith{3}'= 0.01080^{+0.00062}_{-0.00060}$ & \\
$\AmvecWith{3}' = \AmAmpWith{3}' ( \xhat \cos 3 \AmAngWith{3}
			         +\yhat \sin 3 \AmAngWith{3} ) $
& $\begin{array}[t]{@{}c@{}l@{}} 3\AmAngWith{3} & =2.678^{+0.057}_{-0.055} \\ 
 		       \AmAngWith{3} & = 81.15^{+1.09}_{-1.04} \end{array}$

& $\begin{array}[t]{@{}l} \mbox{radians N of W} \\
		       \mbox{degrees E of N} \end{array}$
\\ \tableline
external $m=4$ multipole: &$\AmAmpWith{4}'= 0.00415^{+0.00039}_{-0.00038}$ & \\
$\AmvecWith{4}' = \AmAmpWith{4}' ( \xhat \cos 4 \AmAngWith{4}
			         +\yhat \sin 4 \AmAngWith{4} ) $
& $\begin{array}[t]{@{}c@{}l@{}} 4\AmAngWith{4} & =2.020^{+0.071}_{-0.069} 
                                                                      \,\, \\ 
 		       \AmAngWith{4} & = 28.93^{+1.01}_{-0.99}  \\
 		45^\circ + \AmAngWith{4} & = 73.93 ^{+1.01}_{-0.99} \end{array}$
& $\begin{array}[t]{@{}l} \mbox{radians N of W} \\
		       \mbox{degrees E of N} \\
		       \mbox{degrees E of N} \end{array}$ \\
\tableline
\tableline
undeflected source positions: & ${\bf s}_{p}\cdot\hat{{\bf x}} = 0.8795$ & arcseconds \\
                              & ${\bf s}_{p}\cdot\hat{{\bf y}} = 0.7549$ & arcseconds \\
                              & ${\bf s}_{q}\cdot\hat{{\bf x}} = 0.8815$ & arcseconds \\
                              & ${\bf s}_{q}\cdot\hat{{\bf y}} = 0.7559$ & arcseconds \\
                              & ${\bf s}_{r}\cdot\hat{{\bf x}} = 0.8928$ & arcseconds \\
                              & ${\bf s}_{r}\cdot\hat{{\bf y}} = 0.7626$ & arcseconds \\
                              & ${\bf s}_{s}\cdot\hat{{\bf x}} = 0.8687$ & arcseconds \\
                              & ${\bf s}_{s}\cdot\hat{{\bf y}} = 0.7509$ & arcseconds \\
\tableline
\end{tabular}
\renewcommand{\arraystretch}{1.0}
\end{center}
\caption{
Best-fit model parameters for the deflector model $+\AmvecWith{3}+\AmvecWith{4}$.
See Figure~5
for a graphical depiction of these model parameters. The limits provided for each parameter
are formal 68.3\% confidence limits, computed as described in section~4.1.
For completeness, the lower half of the table lists the undeflected
source positions for the model sources; these are a by-product
of modeling the deflector gravitational potential, and no
confidence ranges on the source positions were calculated.
The image magnifications predicted by this
best-fit model are listed in Table~8.
}
\label{tbl:best-fit-parameters}
\end{table}

\begin{table}
\noindent
\begin{center}
\begin{tabular}{ccc} \tableline
Image & Component & Magnification \\
\tableline \tableline
A1	&	p	&	12.4	\\
A1	&	q	&	12.6	\\
A1	&	r	&	15.3	\\
A1	&	s	&	12.2	\\
A2	&	p	&	$-14.1$	\\
A2	&	q	&	$-14.2$	\\
A2	&	r	&	$-16.4$	\\
A2	&	s	&	$-14.5$	\\
B	&	p	&	5.14	\\
B	&	q	&	5.06	\\
B	&	r	&	4.66	\\
B	&	s	&	5.58	\\
C	&	p	&	$-1.84$	\\
C	&	q	&	$-1.85$	\\
C	&	r	&	$-1.89$	\\
C	&	s	&	$-1.78$	\\
\tableline
\end{tabular}
\end{center}
\caption{
Image magnifications, as predicted by the best-fit
model $+\AmvecWith{3}+\AmvecWith{4}$, which is
specified by the parameters in Table~7.
}
\label{tbl:magnifications}
\end{table}

\begin{table}
\noindent
\begin{center}
\begin{tabular}{lrrrr} \tableline
Model &$\Delta\tau'_{A1p A2p} \times 10^{12}$ & $\Delta\tau'_{A1p Bp} \times 10^{12}$ & $\Delta\tau'_{A1p Cp} \times 10^{12}$ &$\Delta\tau'_{Bp Cp} \times 10^{12}$ \\ \tableline \tableline 
$+\AmvecWith{3}$                              & $0.0519^{+0.0009}_{-0.0009}$ & $-2.326^{+0.038}_{-0.039}$ & $8.66^{+0.09}_{-0.09}$   & $10.99^{+0.12}_{-0.12}$   \\ 
$+\MsummvecWith{3}$                         & $0.0588^{+0.0009}_{-0.0009}$ & $-2.481^{+0.037}_{-0.037}$ & $9.96^{+0.12}_{-0.12}$   & $12.44^{+0.16}_{-0.16}$   \\ 
\tableline 																															     
$+\MsummvecWith{3}+\AmvecWith{4}$             & $0.0660^{+0.0014}_{-0.0013}$ & $-2.616^{+0.047}_{-0.047}$ & $9.87^{+0.32}_{-0.30}$   & $12.49^{+0.31}_{-0.29}$   \\ 
$+\MsummvecWith{3}+\Gvec_{22}$              & $0.0619^{+0.0016}_{-0.0013}$ & $-2.610^{+0.066}_{-0.077}$ & $11.33^{+0.20}_{-0.17}$  & $13.94^{+0.17}_{-0.16}$   \\ 
$+\MsummvecWith{3}+\Gvec_{12}$              & $0.0485^{+0.0012}_{-0.0012}$ & $-2.061^{+0.059}_{-0.068}$ & $8.38^{+0.27}_{-0.23}$   & $10.44^{+0.34}_{-0.29}$   \\ 
\tableline 																															     
$+\AmvecWith{3}+\BmvecWith{3}$                  & $0.0589^{+0.0012}_{-0.0011}$ & $-2.486^{+0.034}_{-0.028}$ & $10.69^{+0.60}_{-0.51}$  & $13.18^{+0.61}_{-0.52}$   \\ 
$+\AmvecWith{3}+\Gvec_{22}$                   & $0.0536^{+0.0003}_{-0.0004}$ & $-1.985^{+0.019}_{-0.020}$ & $8.11^{+0.15}_{-0.15}$   & $10.10^{+0.17}_{-0.17}$   \\ 
$+\AmvecWith{3}+\AmvecWith{4}$                  & $0.0515^{+0.0019}_{-0.0017}$ & $-1.802^{+0.034}_{-0.036}$ & $10.44^{+0.18}_{-0.17}$  & $12.25^{+0.17}_{-0.17}$   \\ 
\tableline 																															     
$+\MsummvecWith{3}+\AmvecWith{4}+\ftWith{3}$  & $0.0680^{+0.0172}_{-0.0072}$ & $-2.694^{+0.273}_{-0.703}$ & $10.10^{+2.49}_{-0.92}$  & $12.80^{+3.18}_{-1.17}$   \\  
$+\MsummvecWith{3}+\Gvec_{22}+\ftWith{3}$   & $0.0709^{+0.0531}_{-0.0174}$ & $-2.912^{+0.664}_{-2.092}$ & $8.64^{+5.48}_{-1.64}$   & $11.55^{+7.57}_{-2.30}$   \\  
$+\MsummvecWith{3}+\Gvec_{12}+\ftWith{3}$   & $0.0525^{+0.0017}_{-0.0020}$ & $-2.202^{+0.075}_{-0.044}$ & $8.98^{+0.16}_{-0.31}$   & $11.18^{+0.21}_{-0.38}$   \\  
\tableline 																															     
$+\AmvecWith{3}+\BmvecWith{3}+\ftWith{3}$       & $0.0493^{+0.0111}_{-0.0051}$ & $-2.133^{+0.161}_{-0.410}$ & $9.49^{+1.48}_{-0.93}$   & $11.62^{+1.86}_{-1.08}$   \\ 
$+\AmvecWith{3}+\Gvec_{22}+\ftWith{3}$        & $0.0588^{+0.0047}_{-0.0051}$ & $-2.183^{+0.194}_{-0.174}$ & $8.57^{+0.63}_{-0.63}$   & $10.75^{+0.77}_{-0.79}$   \\ 
$+\AmvecWith{3}+\AmvecWith{4}+\ftWith{3}$       & $0.0470^{+0.0022}_{-0.0017}$ & $-1.734^{+0.029}_{-0.029}$ & $9.13^{+0.61}_{-0.46}$   & $10.86^{+0.62}_{-0.45}$   \\ 
\tableline
\end{tabular}
\end{center}
\caption[Dimensionless time delays for models fitted to MG~J0414+0534 ]
{
Dimensionless time delays for models fitted to MG~J0414+0534, using the
stat-width estimates of the observational uncertainties.  The
dimensionless time delay is defined in equation~64.
The results are reported for the brightest component,~p.
Variations would first be observed in image B, then A1, A2, and finally C.
}
\label{tbl:dimensionless-time-delays}
\end{table}

\begin{table}
\noindent
\begin{center}
\begin{tabular}{ccc} \tableline
$\Omega_{m}$ & $\Omega_{\Lambda}$ & $\Delta t / \Delta\tau \times 10^{-12}$ \\
             &                    & ($h^{-1}_{75}$ days) \\
\tableline \tableline
1 & 0 & 6.794 \\
0.3 & 0.7 & 7.654 \\
0 & 1 & 7.166 \\
0 & 0 & 8.785 \\
\tableline 
\end{tabular}
\end{center}
\caption{
Values of the conversion factor between the time delay and the
dimensionless time delay, in various cosmological scenarios.
}
\label{tbl:conversion-factors}
\end{table}

\begin{table}
\noindent
\begin{center}
\renewcommand{\arraystretch}{1.5}
\begin{tabular}{lcc} \tableline
Model &$\Delta\tau_{Bp A1p}/\Delta\tau_{A1p Cp}$ & $\Delta\tau_{Bp A2p}/\Delta\tau_{A2p Cp}$ \\ \tableline \tableline
$+\AmvecWith{3}$                       & $0.268^{+0.003}_{-0.003}$ & $0.276^{+0.003}_{-0.003}$  \\  
$+\MsummvecWith{3}$                  & $0.249^{+0.002}_{-0.002}$ & $0.256^{+0.002}_{-0.002}$  \\  
\tableline 
$+\MsummvecWith{3}+\AmvecWith{4}$      & $0.265^{+0.010}_{-0.011}$ & $0.274^{+0.011}_{-0.011}$  \\  
$+\MsummvecWith{3}+\Gvec_{22}$       & $0.230^{+0.007}_{-0.008}$ & $0.237^{+0.007}_{-0.008}$  \\  
$+\MsummvecWith{3}+\Gvec_{12}$       & $0.246^{+0.002}_{-0.002}$ & $0.253^{+0.002}_{-0.002}$  \\  
\tableline 
$+\AmvecWith{3}+\BmvecWith{3}$           & $0.233^{+0.011}_{-0.011}$ & $0.239^{+0.011}_{-0.012}$  \\  
$+\AmvecWith{3}+\Gvec_{22}$            & $0.245^{+0.004}_{-0.004}$ & $0.253^{+0.004}_{-0.004}$  \\  
$+\AmvecWith{3}+\AmvecWith{4}$           & $0.173^{+0.006}_{-0.006}$ & $0.178^{+0.006}_{-0.006}$  \\  
\tableline 
$+\MsummvecWith{3}+\AmvecWith{4}+\ftWith{3}$      & $0.267^{+0.015}_{-0.012}$ & $0.275^{+0.015}_{-0.013}$  \\  
$+\MsummvecWith{3}+\Gvec_{22}+\ftWith{3}$       & $0.337^{+0.020}_{-0.017}$ & $0.348^{+0.021}_{-0.018}$  \\  
$+\MsummvecWith{3}+\Gvec_{12}+\ftWith{3}$       & $0.245^{+0.001}_{-0.001}$ & $0.253^{+0.001}_{-0.001}$  \\  
\tableline 
$+\AmvecWith{3}+\BmvecWith{3}+\ftWith{3}$  & $0.225^{+0.013}_{-0.008}$ & $0.231^{+0.014}_{-0.008}$  \\  
$+\AmvecWith{3}+\Gvec_{22}+\ftWith{3}$   & $0.255^{+0.016}_{-0.014}$ & $0.263^{+0.016}_{-0.015}$  \\  
$+\AmvecWith{3}+\AmvecWith{4}+\ftWith{3}$  & $0.190^{+0.012}_{-0.012}$ & $0.196^{+0.012}_{-0.012}$  \\  
\tableline
\end{tabular}
\renewcommand{\arraystretch}{1.0}
\end{center}
\caption{
Time delay ratios for models fitted to MG~J0414+0534. The quoted errors
were computed using the stat-width estimates for the positional errors.
}
\label{tbl:MMT-time-delay-ratios}
\end{table}

\end{document}